\documentclass[preprint,aps,prd,superscriptaddress,
nofootinbib]{revtex4}%
\usepackage{graphicx}
\usepackage{amsmath}
\usepackage{amssymb}
\usepackage{bm}
\usepackage{epsfig}
\usepackage{mathrsfs}
\usepackage{hyperref}

\begin{document}
\preprint{CERN-TH-2016-200}
\preprint{EFI-16-18}
\vspace{-2.5cm}
\title{\mbox{}\\[10pt] Higgs-stoponium mixing near the stop-antistop 
threshold}
\author{Geoffrey~T.~Bodwin}
\email[]{gtb@anl.gov}
\affiliation{High Energy Physics Division, Argonne National Laboratory,
Argonne, Illinois 60439, USA}
\author{Hee~Sok~Chung}
\email[]{hee.sok.chung@cern.ch}
\affiliation{High Energy Physics Division, Argonne National Laboratory,
Argonne, Illinois 60439, USA}
\affiliation{Theory Department, CERN, 1211 Geneva 23, Switzerland}
\author{Carlos E.~M.~Wagner}
\email[]{cwagner@anl.gov}
\affiliation{High Energy Physics Division, Argonne National Laboratory,
Argonne, Illinois 60439, USA}
\affiliation{Enrico Fermi Institute, University of Chicago, Chicago, IL 
60637, USA}
\date{\today}
\vspace{-1.cm}
\begin{abstract}
Supersymmetric extensions of the standard model contain additional heavy
neutral Higgs bosons that are coupled to heavy scalar top quarks
(stops). This system exhibits interesting field theoretic phenomena when
the Higgs mass is close to the stop-antistop production threshold.
Existing work in the literature has examined the digluon-to-diphoton
cross section near threshold and has  focused on enhancements in
the cross section that might arise either from the perturbative
contributions to the  Higgs-to-digluon and Higgs-to-diphoton form factors
or from mixing of the Higgs boson with stoponium states. Near threshold,
enhancements in the relevant amplitudes that go as inverse powers of the
stop-antistop relative velocity require resummations of perturbation
theory and/or nonperturbative treatments.  We present a complete
formulation of  threshold effects at leading order in the stop-antistop
relative velocity in terms of nonrelativistic effective field theory. We
give detailed numerical calculations for the case in which the
stop-antistop Green's function is modeled with a Coulomb-Schr\"odinger
Green's function. We find several general effects that do not appear in
a purely perturbative treatment. Higgs-stop-antistop mixing effects
displace physical masses from the threshold region, thereby rendering
the perturbative threshold enhancements inoperative. In the case of
large Higgs-stop-antistop couplings, the displacement of a physical
state above threshold substantially increases its width, owing to its
decay width to a stop-antistop pair, and greatly reduces its
contribution to the cross section.

\end{abstract}
\maketitle

\section{Introduction}

In extensions of the standard model (SM), new heavy particles typically
appear. For example, supersymmetric extensions of the SM include heavy stop 
quarks (stops), which are the supersymmetric partners of a top quark
\cite{Haber:1984rc,Nilles:1983ge,Drees:2004jm,Baer:2006rs,Martin:1997ns}.
A stop quark and a stop antiquark (antistop) can bind to form a
spectrum of stoponium states.  The decays of these states into two
photons potentially provide a clean signal for their detection. However,
stoponium states typically have rather small gluon-fusion cross sections
times branching ratios into two
photons~\cite{Drees:1993uw,Martin:2008sv,Martin:2009dj,Younkin:2009zn,
Kats:2016kuz,Choudhury:2016jbc,Carena:2016bnq}.

Extensions of the SM can also contain heavy Higgs
bosons~\cite{Branco:2011iw}. The presence of new Higgs doublets is
motivated by weak-scale extensions of the SM that aim
to address the disparity between the electroweak and Planck scales and
to provide explanations of the origins of flavor and of the
matter-antimatter asymmetry. In these theories, the SM description is
recovered in the so-called decoupling regime, in which the masses of the
heavy Higgs bosons become large. In such a regime, the heavy neutral
Higgs bosons may decay into pairs of third-generation fermions,
including top quarks. Owing to the presence of these tree-level
decays, the branching ratio of loop-induced decay processes is
suppressed, making it difficult to observe the decay of the heavy Higgs boson 
to two photons.

The interplay of a heavy-stop-antistop system with a heavy Higgs boson
whose mass is near the stop-antistop production threshold results in
interesting and intricate new phenomena. Loop-induced processes may be
enhanced in the presence of heavy quarks or squarks that are strongly
coupled to the heavy Higgs boson. In supersymmetric extensions of the
SM, the dimensionful coupling of stops to the heavy Higgs bosons is
governed by the Higgsino mass parameter $\mu$. Loop-induced processes
may be significantly modified if the heavy quarks or squarks have masses
that are comparable to that of the heavy Higgs boson. The modification
to loop-induced processes may be especially important if there is a
production threshold for heavy particle-antiparticle pairs that is close
to the Higgs mass~\cite{Drees:1989du}.

Near threshold, QCD perturbation-expansion contributions of relative
order $n$ may receive $1/v^n$ enhancements, where $v$ is half of the
relative velocity between the stop and the antistop in the
stop-antistop center-of-momentum (CM) frame. Such enhanced
contributions are typically proportional to $\alpha_s^n(m_{\tilde t}
v)/v^n$, where $\alpha_s$ is the strong-interaction running coupling and
$m_{\tilde{t}}$ is the stop-quark mass. The presence of these enhanced
contributions requires a resummation of the perturbative
contributions, which, among other things, takes into account the
formation of stop-antistop bound states.

The issue of gluon-fusion Higgs production and decay to diphotons at the
stop-antistop production threshold has been addressed recently in
Ref.~\cite{Djouadi:2016oey}. The authors of Ref.~\cite{Djouadi:2016oey}
recognized the possibility that Higgs-stoponium mixing and the
formation of stop-antistop bound states could have a significant
effect on the $gg\to H \to \gamma\gamma$ rate. They pointed out, as
is stressed in Ref.~\cite{Melnikov:1994jb}, that stoponium effects lead
to Higgs-digluon ($Hgg$) and Higgs-diphoton ($H\gamma\gamma$) form
factors that are enhanced relative to the form factors that are obtained
in fixed-order perturbative calculations.  We note, though, that the
description in Refs.~\cite{Melnikov:1994jb,Djouadi:2016oey} does not
explicitly account for all of the Higgs-stoponium mixing
effects~\cite{Drees:1993uw}.

In this article, we provide a detailed analysis of the interplay between
a heavy Higgs boson and a heavy stop-antistop pair for Higgs masses that
are close to the stop-antistop production threshold.\footnote{ In the
cases of SM extensions that contain more than one stop quark, we will 
assume that the heaviest stop mass eigenstate is significantly heavier
than the lightest one. This implies that the Higgs mass is far below the
threshold for production of a heaviest stop and a heaviest antistop.
Therefore, the heaviest-stop contribution becomes subdominant with
respect to the lightest-stop contribution. We will consider the
heaviest-stop contribution to be a perturbation to the rates that are
computed in this work.  It should be taken into account in precision
studies. For large values of the trilinear Higgs-stop-antistop coupling, the
coupling of the heavy stop to the Higgs boson is of the same order as
and opposite in sign to the coupling of the lightest stop to the Higgs
boson. Hence, the contribution of the heavy stop to the diphoton rate
may be non-negligible in the regime of strong Higgs-stop-antistop
coupling.} We take into account threshold enhancements to the stop-antistop amplitudes
by means of the analogues for scalar quarks of the effective field
theories nonrelativistic QED (NRQED)~\cite{Caswell:1985ui} and
nonrelativistic QCD (NRQCD)~\cite{Lepage:1992tx,Bodwin:1994jh}. In this
framework, we are also able to give a description of Higgs-stop-antistop
mixing near threshold that incorporates fully the effects that are
contained in the stop-antistop Green's function.\footnote{In
Ref.~\cite{Drees:1989du}, the effects of mixing between the Higgs boson
and discrete stoponium resonances were considered.}

In order to make contact with recent numerical work on the
Higgs-stop-antistop system we choose, in our numerical work, a
stop mass of $375$~GeV, and we assume that the Higgs mass is near the
stop-antistop production threshold, which is $750$~GeV. Much of the
recent numerical work in the literature was motivated by initial results
from the ATLAS and CMS experiments that showed excesses in rates for the
process $pp\to \gamma\gamma$ at diphoton masses around 750~GeV
\cite{ATLAS:2016jcu,CMS:2016owr}. (For an extensive list of theoretical
work that is related to this signal, see
Ref.~\cite{Strumia:2016wys}.) However, we emphasize that our work is not
tied to any particular phenomenological model and that it is aimed at
understanding the general features of a Higgs-stop-antistop system near
threshold.

We find that effects of Higgs-stop-antistop mixing go beyond the
modification of the $Hgg$ and $H\gamma\gamma$ form factors and
significantly change the diphoton production rate near the stop-antistop
threshold with respect to the rate that would be obtained from
the simple addition of the Higgs and stoponium contributions. For Higgs
masses near threshold, we find that several mechanisms that arise from
Higgs-stop-antistop mixing suppress the diphoton rate relative to the
rates that are obtained in perturbative calculations: (1) for
small stop widths, mixing significantly increases the width of the
narrowest physical state relative to the width of the unmixed stoponium
state; (2) mixing shifts masses of physical states away from the region in
which form-factor enhancements occur; (3) mixing shifts some physical-state
masses above threshold, where, in the case of strong
Higgs-stop-antistop couplings,  the states develop large decay widths
into stop-antistop pairs; (4) for strong Higgs-stop-antistop couplings,
the mixing changes the  heights and widths of the physical resonances
that lie below threshold.

In some of our numerical work, we employ rather large values of the
Higgs-stop-antistop coupling. In some specific models of the
Higgs-stop-antistop system, such large values of the coupling could lead
to the presence of color-symmetry-breaking minima in the potential for
the vacua~\cite{Casas:1995pd,Blinov:2013fta,Chowdhury:2013dka}.
Since, in our work, we are not focused on any specific model realization
of the Higgs-stop-antistop system, we will not study these possible
constraints. However, they would have to be taken into account in the
construction of detailed models.

The remainder of this article is organized as follows. In
Sec.~\ref{sec:EFT}, we describe the  effective-field-theory
approach that we employ. We also give formulas for the
$gg\to\gamma\gamma$ amplitude that account fully for the
Higgs-stop-antistop mixing. Section~\ref{sec:breit-wigner} contains a 
discussion of a simplified model in which the stop-antistop states are 
replaced by a single Breit-Wigner resonance. That model exhibits a 
number of the features of the full theory. 
In Sec.~\ref{sec:numerical}, we present and
discuss our numerical results for the $gg\to\gamma\gamma$ amplitude and
cross section as a function of the Higgs mass, relative to the
stop-antistop threshold. Section~\ref{sec:conclusions} contains our
conclusions.

\section{Effective-field-theory approach\label{sec:EFT}}

\subsection{NRQED/NRQCD analogues for scalar quarks}

We wish to compute the amplitude for $gg\to  \gamma\gamma$ in the
presence of a Higgs boson with mass $m_H$ that couples to a heavy stop
quark and a heavy antistop quark, each of which have mass $m_{\tilde
t}$. We are concerned with the situation in which $m_H$ is near $2
m_{\tilde t}$, the threshold for stop-antistop production. Because the
amplitude is computed near threshold, there can be important effects
from the binding or near binding of the stop-antistop pair that are not
captured in fixed-order perturbation theory. It is convenient to take
these effects into account by making use of the analogues for scalar
quarks of the effective field theories NRQED
\cite{Caswell:1985ui} and NRQCD
\cite{Lepage:1992tx,Bodwin:1994jh}.\footnote{In Ref.~\cite{Kim:2014yaa}, 
an effective field theory for stoponium systems was developed, and 
a resummation of threshold logarithms in the stoponium production cross 
section was carried out by making use of soft-collinear effective 
theory~\cite{Bauer:2000yr}. However, the formalism in Ref.~\cite{Kim:2014yaa} 
does not address the possibility of stoponium-Higgs mixing.}

We will carry out the effective-field-theory computation at the leading
nontrivial order in the heavy-stop velocity $v$ in the stop-antistop
CM frame, where $v$ is given by
\begin{equation}
v=|{\bm p}|/m_{\tilde t}=\sqrt{2m_{\tilde t}E}/m_{\tilde t}.
\end{equation}
Here ${\bm p}$ is the 3-momentum of the stop in the stop-antistop (or
$\gamma\gamma$) CM frame,
\begin{equation}
E=\sqrt{\hat s}-2m_{\tilde t}
\end{equation}
is the nonrelativistic CM energy of the stop-antistop system, 
$\sqrt{\hat s}$ is the partonic CM energy,
\begin{equation}
\sqrt{\hat s}=m_{\gamma\gamma},
\end{equation}
and $m_{\gamma\gamma}$ is the $\gamma\gamma$ mass. 

The effective field theory is an expansion in powers of $v$. Hence, our
calculation should be valid as long as $v$ is much less than 1. We
expect corrections to our calculation of the $gg\to \gamma\gamma$
amplitude to be of relative order $v^2$.

The heavy-stop part of the effective Lagrangian density that we will use,
which is valid at the leading order in $v$, is 
\begin{eqnarray}\label{heavy-action}
{\cal L}_{\tilde t\tilde t}&=&
\psi^\dagger\biggl(2im_{\tilde t}D_0+{\bm D}^2\biggl)\psi +
\chi^\dagger\biggl(2im_{\tilde t}D_0+{\bm D}^2\biggl)\chi 
-iC_{H\tilde t\tilde t}H(\psi^\dagger\chi+\chi^\dagger\psi)\nonumber\\
&&+(i/2)C_{gg\tilde t\tilde t}\frac{1}{N_c^2-1}
(\psi^\dagger\chi+\chi^\dagger\psi)G_{\mu\nu}^a
G^{a\mu\nu}
+(i/2)C_{\gamma\gamma\tilde t\tilde t}(\psi^\dagger\chi+\chi^\dagger\psi)
F_{\mu\nu} F^{\mu\nu}\nonumber\\
&&+(i/2)C_{ggH}\frac{1}{N_c^2-1}HG_{\mu\nu}^a G^{a\mu\nu}
+(i/2)C_{\gamma\gamma H}HF_{\mu\nu} F^{\mu\nu}\nonumber\\
&&-iC_{\tilde t\tilde t H \tilde t\tilde t}
\frac{1}{N_c}\psi^\dagger\chi\chi^\dagger\psi
+i {\rm Im}\, T_{\tilde t\tilde t\to gg\to \tilde t\tilde t}
\frac{1}{N_c}\psi^\dagger\chi\chi^\dagger\psi,
\end{eqnarray}
where $m_{\tilde t}$ is the stop pole mass, $\psi$ is the field that
annihilates a stop, $\chi$ is the field that creates an antistop,
$G_\mu^a$ is the gluon field with adjoint color index $a$,
$G_{\mu\nu}^a$ is the gluon field strength with adjoint color index $a$,
$A_\mu$ is the electromagnetic field, and $F_{\mu\nu}$ is the
electromagnetic field strength. The covariant derivative contains both
the electromagnetic field and the gluon field:
\begin{subequations}%
\begin{equation}
D_\mu=\partial_\mu-iee_{\tilde t}A_\mu-ig G_\mu^a t_a,
\end{equation}
where $e$ is the electromagnetic coupling, $e_{\tilde t}$ is the
stop-quark charge, $g$ is the strong-interaction coupling, and $t_a$ is
an $SU(3)$ matrix in the adjoint representation that is normalized to
\begin{equation}
{\rm Tr}\, t_a t_b=T_R \delta_{ab}=(1/2)\delta_{ab}.
\end{equation}
\end{subequations}%
In the calculations in this paper, we ignore the couplings of stops and
antistops to the electromagnetic field, except in the annihilation of a
stop-antistop pair into two photons. The $C_i$'s are short-distance
coefficients, which will be determined by matching the effective
theory with full QED and
full QCD at the stop-antistop threshold. The short-distance
coefficient $C_{\tilde t\tilde t H \tilde t\tilde t}$ takes into account
the $t$-channel exchange of the heavy Higgs boson between the stop and
the antistop. The quantity ${\rm Im} T_{\tilde t\tilde t\to gg\to
\tilde t\tilde t}$ is also a short-distance coefficient that accounts
for decays of a stop-antistop pair into two gluons. It is given by
($-2$) times the imaginary part of the contribution to the stop-antistop
forward $T$-matrix that contains a two-gluon intermediate state,
evaluated at the stop-antistop threshold. Note that, because we are
working at leading order in $v$, there are only $S$-wave couplings to
the stop-antistop pairs.

\subsection{Computation of the short-distance coefficients}

In this section, we compute the short-distance coefficients in
Eq.~(\ref{heavy-action}) by matching the effective theory to the full
theory. Because our focus is on the formulation of the calculation and
on the qualitative features of the threshold physics, we work at the
lowest nontrivial order in the electromagnetic and strong couplings.
Therefore, one should take care in comparing our numerical results with
those in the literature, which often are performed at next-to-leading
order, and, therefore, include two-loop effects in the
couplings of the Higgs boson to digluons and
diphotons~\cite{Spira:1995rr,Harlander:2012pb,Bagnaschi:2014zla,Dittmaier:2014sva}.\footnote{\label{footnote:alpha-expansion} An expansion of in
powers of $\alpha_s$ is valid for the short-distance coefficients, since
they contain no $1/v$ enhancements.  However, in the computation of the
effective-field-theory amplitudes in Sec.~\ref{sec:comp-amplitudes}, the
expansion in powers of $\alpha_s$ can fail because there are
contributions to the stop-antistop Green's function in order
$\alpha_s^n$ that are enhanced by factors $1/v^n$. We compute these
contributions to all orders in $\alpha_s$.}

We compute the short-distance coefficients that appear at 
the Born level by evaluating the corresponding amplitude in the full 
theory at the stop-antistop threshold. We compute the short-distance 
coefficients that appear at one-loop level by evaluating the one-loop 
amplitude in the full theory at threshold and subtracting the 
corresponding one-loop amplitude in the effective theory.

The short-distance coefficients $C_{gg\tilde t\tilde t}$,
$C_{\gamma\gamma\tilde t\tilde t}$, and $C_{\tilde t\tilde t H \tilde
t\tilde t}$ are easily obtained by carrying out Born-level
calculations in full QCD at the stop antistop threshold. They are given
by 
\begin{subequations}%
\begin{eqnarray}
C_{gg\tilde t\tilde t}&=&8i\pi\alpha_s(m_{\tilde t}) \frac{T_R}{\sqrt{N_c}},\\
C_{\gamma\gamma\tilde t\tilde t}&=&8i\pi\alpha e_{\tilde t}^2\sqrt{N_c},\\
C_{\tilde t\tilde t H \tilde t\tilde t}&=&iN_c\frac{g_{H\tilde t\tilde 
t}^2}{4 m_{\tilde t}^2},
\end{eqnarray}
\end{subequations}%
where $\alpha=e^2/(4\pi)$, $e_{\tilde t}$ is the stop electromagnetic
charge, $\alpha_s=g^2/(4\pi)$, and $N_c=3$ is the number of $SU(3)$
colors. We have chosen the effective-field-theory factorization
scale to be $m_{\tilde t}$, which accounts for the argument of
$\alpha_s$. We have also taken the color-singlet projection of the
stop-antistop pairs, making use of the projector
\begin{equation}
P_{ij}=\delta_{ij}/\sqrt{N_c},
\label{color-sing-proj}
\end{equation} 
where $i$ and $j$ are the squark and antiquark color indices,
respectively.  

The short-distance coefficient $C_{H\tilde t\tilde
t}$ is simply the Born-level Higgs-stop-antistop coupling, $g_{H
\tilde{t} \tilde{t}}$, rescaled by a factor of $\sqrt{N_c}$:
\begin{equation}
C_{H\tilde t\tilde t}=i \sqrt{N_c} g_{H \tilde{t} \tilde{t}} \equiv - i \sqrt{N_c} \kappa m_{\tilde t},
\end{equation}
where we have normalized the Higgs coupling to stops in terms of the 
stop mass, with $\kappa$ being an adjustable parameter.

The short-distance coefficients $C_{ggH}$ and $C_{H\gamma\gamma}$ are 
generated by quark loops and stop loops. We take into account the 
$b$-quark, $t$-quark, and stop loops, which give the most important
contributions. In full QCD, we have the
amplitude~\cite{Wilczek:1977zn,Ellis:1979jy,Gunion:1989we}

\begin{equation}
i {\cal M}_{gg \to H}(\hat s) =
\biggl(\epsilon_1 \cdot \epsilon_2-\frac{\epsilon_1 
\cdot k_2 \epsilon_2 \cdot k_1}{k_1\cdot k_2}\biggr)\delta_{ab}
{\cal A}_{gg\to H}(\hat s),
\label{ggtoH}
\end{equation}

where 
\begin{equation}
{\cal A}_{gg\to H} (\hat s)
= \frac{i \alpha_s}{8 \pi} T_F \hat{s}  
\left[ \frac{2 g_{Hbb}}{m_b} A_{1/2} (\tau_b) +
\frac{2 g_{Htt}}{m_t}  A_{1/2} (\tau_t) +
\frac{g_{H \tilde t \tilde t}}{m_{\tilde t}^2}  A_0 (\tau_{\tilde t}) \right],
\end{equation}
\begin{subequations}%
\begin{eqnarray}
A_{1/2} (\tau) &=& 2 [ \tau + \tau (1-\tau) f(\tau) ]
,
\\
A_0 (\tau) &=& -\tau [1-\tau f(\tau)],
\end{eqnarray}
\end{subequations}%
\begin{equation}
f(\tau) = \begin{cases}
\arcsin^2 (1/\sqrt{\tau}) & \tau \ge 1,
\\
-\frac{1}{4} \left[ \log \left( 
\tfrac{1+\sqrt{1-\tau}}{1-\sqrt{1-\tau}}
\right) - i \pi \right]^2 & \tau <1,
\end{cases}
\end{equation}
and
\begin{subequations}%
\begin{eqnarray}
\tau_b &=& \frac{4 m_b^2}{\hat s},\\
\tau_t &=& \frac{4 m_t^2}{\hat s},\\
\tau_{\tilde t} &=& \frac{4 m_{\tilde t}^2}{\hat s}.
\end{eqnarray}
\end{subequations}%
Here, $(k_1,\epsilon_1)$ and $(k_2,\epsilon_2)$ are the
$(\rm{momentum},{\rm polarization})$ of the initial gluons, $a$ and $b$
are the gluon color indices, $m_b$ and $e_b$ are the bottom-quark mass
and electric charge, $m_t$ and $e_t$ are the top-quark mass and electric
charge,
\begin{subequations}%
\begin{eqnarray}
g_{Hbb}&=&\frac{g_{\rm EW} m_b}{2 m_W} \tan\beta,\\
g_{Htt}&=&- \frac{g_{\rm EW} m_t}{2 m_W \tan\beta} ,\\
g_{H\tilde t\tilde t}&=&-\kappa m_{\tilde t},
\end{eqnarray}
\end{subequations}%
$g_{\rm EW}$ is the electroweak coupling, $m_W$ is the $W$-boson mass,
$\tan\beta$ is the  ratio of heavy and light Higgs vacuum expectation values in
a supersymmetric model, 
and we have listed the values of the couplings in the heavy-Higgs-boson
decoupling limit, ignoring small deviations of the couplings from those
values.

Similarly, in the case of $H\to \gamma\gamma$, we have the
amplitude~\cite{Ellis:1975ap,Shifman:1979eb,Gunion:1989we,Djouadi:1993ji}

\begin{equation}
i {\cal M}_{H\to\gamma\gamma}(\hat s) =
\biggl( \epsilon_3 \cdot \epsilon_4 -\frac{
\epsilon_3 \cdot k_4 \epsilon_4 \cdot k_3}{k_3 \cdot k_4}\biggr) 
{\cal A}_{H\to\gamma\gamma}(\hat s),
\label{gammagammatoH}
\end{equation}  

where
\begin{equation}
{\cal A}_{H\to \gamma\gamma} (\hat s)
= \frac{i \alpha}{8 \pi} N_c \hat{s}
\left[ \frac{2 g_{Hbb}}{m_b} e_b^2 A_{1/2} (\tau_b) +
\frac{2 g_{Htt}}{m_t} e_{\tilde t}^2 A_{1/2} (\tau_t) +
\frac{g_{H \tilde t \tilde t}}{m_{\tilde t}^2} e_t^2 A_0 (\tau) \right],
\end{equation}
and $(k_3,\epsilon_3)$ and $(k_4,\epsilon_4)$ are the 
$({\rm momentum},{\rm polarization})$ of the final photons.

The corresponding quantities in the effective theory are produced by the 
stop loop that is generated by 
the $C_{H\tilde t\tilde t}$, $C_{gg\tilde t\tilde t}$, and 
$C_{\gamma\gamma\tilde t\tilde t}$ terms in the effective action.
In the modified-minimal-subtraction ($\overline{\rm MS}$) scheme, we obtain
\begin{subequations}\label{eff-th-gg-gamgam-H}%
\begin{eqnarray}
{\cal A}_{gg\to H}^{\rm eff}=C_{gg\tilde t\tilde t}
C_{H\tilde t\tilde t}
\frac{-i}{16\pi m_{\tilde t}}\sqrt{-m_{\tilde t}(E+i\epsilon)},\\
{\cal A}_{H\to \gamma\gamma}^{\rm eff}=C_{\gamma\gamma\tilde t\tilde t}
C_{H\tilde t\tilde t}
\frac{-i}{16\pi m_{\tilde t}}\sqrt{-m_{\tilde t}(E+i\epsilon)}.
\end{eqnarray}
\end{subequations}%

The effective-theory amplitudes in Eq.~(\ref{eff-th-gg-gamgam-H}) vanish
at the stop-antistop threshold ($E=0$). Therefore, the short-distance
coefficients $C_{ggH}$ and $C_{\gamma\gamma H}$ are obtained simply by
evaluating ${\cal A}_{gg \to H}$ and $ {\cal A}_{H\to \gamma\gamma}$ at
stop-antistop threshold:
\begin{subequations}%
\begin{eqnarray}
C_{ggH}&=&{\cal A}_{gg \to H}(4m_{\tilde t}^2),\\
C_{\gamma\gamma H}&=&{\cal A}_{H\to \gamma\gamma}(4m_{\tilde t}^2).
\end{eqnarray}
\end{subequations}%

Finally, as we have mentioned, the short-distance coefficient ${\rm Im}
T_{\tilde t\tilde t\to gg\to \tilde t\tilde t}$ is obtained by computing
the contribution from a two-gluon intermediate state to the
imaginary part of the stop-antistop forward $T$-matrix, evaluated at the
stop-antistop threshold. By making use of unitarity, one can obtain this
quantity simply from a cut diagram. The result is
\begin{equation}
2{\rm Im} T_{\tilde t\tilde t\to gg\to \tilde t\tilde t}=
\frac{1}{\pi}|C_{gg \tilde t \tilde t}|^2.
\end{equation}

\subsection{Computation of the $ gg\to \gamma\gamma$ 
amplitude\label{sec:comp-amplitudes}}

In the nonrelativistic effective theory, the stop-antistop
interactions can be taken into account by considering the stop-antistop
Green's function
\begin{eqnarray}
G_{\tilde t \tilde t}(\hat s)&=&P_{ij}P_{kl}\int dx_0\,
e^{i(\sqrt{\hat s}-2 m_{\tilde t\tilde t}) x_0}
\langle 0|\chi_i^\dagger(x_0,{\bf 0}) 
\psi_j(x_0,{\bf 0})\psi_k^\dagger(0,{\bf
0})\chi_l(0,{\bf 0})|0\rangle.
\end{eqnarray}
Note that
the fields in the Green's function are evaluated at zero spatial
separation and that color-singlet projections of the initial stop and
antistop and the final stop and antistop have been taken by making use
of the projectors $P_{ij} P_{kl}$ [Eq.~(\ref{color-sing-proj})]. The 
Green's function contains all of the effects of the $1/v^n$ enhancements 
that we have mentioned.

The Green's function $G_{\tilde t \tilde t}(\hat s)$ can be evaluated
in a systematic expansion in powers of $v$ by considering a
reformulation of the nonrelativistic effective theory in terms of an
effective theory that is the scalar-squark analogue of potential NRQCD
\cite{Brambilla:1999xf}. In this effective theory, the Green's 
function can be evaluated by considering interactions of the stop and antistop
through nonrelativistic potentials. (The potentials scale with definite
powers of $v$.) The $1/v^n$ enhancements arise from $n$ exchanges
of heavy-stop-antistop potentials.  A resummation of the 
potential exchanges, through the use of the Schr\"odinger equation,
brings the $1/v^n$ enhancements under control. The resummation of the
potential exchanges takes into account, among other things, the
formation of stop-antistop bound states. The potentials incorporate both
perturbative and nonperturbative effects.  If $m_{\tilde t}v$ is
sufficiently large, they are well approximated by perturbative
expressions, but they are also valid when $m_{\tilde t}v$ is in the
nonperturbative regime. 

We can take into account the four-fermion terms in the effective action 
that are proportional to $C_{\tilde t\tilde t H \tilde t\tilde t}$ and 
${\rm Im} T_{\tilde t\tilde t\to gg\to \tilde t\tilde t}$
by replacing $G_{\tilde t \tilde t}(\hat s)$ with \begin{equation}
\tilde{G}_{\tilde t \tilde t}(\hat s)=
\frac{1}
{G_{\tilde t \tilde t}^{-1}(\hat s)(0,0,E)
-C_{\tilde t\tilde t H \tilde t\tilde t}
+{\rm Im} T_{\tilde t\tilde 
t\to gg\to \tilde t\tilde t}}.
\end{equation}
In the case of a stoponium state, this replacement accounts for the
decay width into two gluons, which, for small values of the stop width,
is the dominant stoponium decay width.\footnote{ For large values of the
coupling of the SM-like Higgs boson to a stop-antistop pair, the
stoponium state can also decay with a significant rate into a pair
of 125~GeV Higgs bosons. This occurs, for instance, in the minimal
supersymmetric standard model, for
large values of the stop mixing parameters~\cite{Dreiner:2016wwk}. In
our work, we have assumed that this coupling takes moderate values
and, consequently, that the decay width of the stoponium state
into pairs of SM-like Higgs bosons is much smaller than its decay width
into gluon pairs.}

The coupling of the Higgs-boson to a stop-antistop pair leads to
several contributions to the $gg\to \gamma\gamma$ amplitude. 
We write these contributions to the amplitude as
\begin{equation}
i{\cal M}(gg\to \gamma\gamma)_i=
\biggl(\epsilon_1 \cdot \epsilon_2-\frac{\epsilon_1 
\cdot k_2 \epsilon_2 \cdot k_1}{k_1\cdot k_2}\biggr)
\biggl( \epsilon_3 \cdot \epsilon_4 -\frac{
\epsilon_3 \cdot k_4 \epsilon_4 \cdot k_3}{k_3 \cdot 
k_4}\biggr)\delta_{ab} A_i(gg\to \gamma\gamma).
\end{equation}

There is a contribution in which the initial $gg$ pair transitions to
the Higgs boson and the Higgs boson transitions to a $\gamma\gamma$ pair:
\begin{subequations}\label{A1}%
\begin{eqnarray}
A_1(gg\to \gamma\gamma)&=&C_{ggH}\biggl[
S_H(\hat s)+S_H(\hat s)
C_{H\tilde t\tilde t}\tilde{G}_{\tilde t\tilde t}(\hat s)C_{H\tilde t\tilde t}
S_H(\hat s) + \ldots\biggr] C_{\gamma\gamma H}\label{A1a}\\
&=&C_{gg H} 
\frac{1}{S_H^{-1}(\hat s)-C_{H\tilde t\tilde t}^2 \tilde{G}_{\tilde t\tilde t}
(\hat s)}
C_{\gamma\gamma H}\label{A1b}\\
&=&C_{gg H}
\frac{i}{\hat s-m_H^2+im_H\Gamma_H-iC_{H\tilde t\tilde t}^2
\tilde{G}_{\tilde t\tilde t}(\hat s)}C_{\gamma\gamma H}.
\end{eqnarray}
\end{subequations}%
Here, 
\begin{equation}
S_H=\frac{i}{\hat s-m_H^2+im_H\Gamma_H}
\end{equation}
is the Higgs propagator (that is, the Higgs Green's function in the
absence of Higgs-stop-antistop interactions), $m_H$ is the Higgs pole
mass, and $\Gamma_H$ is the Higgs width.

There is  also a contribution in which the initial $gg$ state
transitions to a Higgs boson, which transitions to a stop-antistop pair,
which transitions to a $\gamma\gamma$ pair:
\begin{eqnarray}
A_2(gg\to \gamma\gamma)
&=&C_{ggH}\frac{i}{\hat s-m_H^2+im_H\Gamma_H-iC_{H\tilde t\tilde t}^2
\tilde{G}_{\tilde t\tilde t}(\hat s)}
C_{H\tilde t\tilde t} \tilde{G}_{\tilde t\tilde t}(\hat s)
C_{\gamma\gamma\tilde t\tilde t}.
\end{eqnarray}

There is a contribution in which the initial $gg$ pair transitions 
to a stop-antistop pair, which transitions to a Higgs boson, which 
transitions to a $\gamma\gamma$ pair:
\begin{eqnarray}
A_3(gg\to \gamma\gamma)
&=&C_{gg\tilde t\tilde t} \tilde{G}_{\tilde t\tilde t}(\hat s) C_{H\tilde t\tilde t}
\frac{i}{\hat s-m_H^2+im_H\Gamma_H-iC_{H\tilde t\tilde t}^2
\tilde{G}_{\tilde t\tilde t}(\hat s)}
C_{\gamma\gamma H}.
\end{eqnarray}

There is a contribution in which the initial $gg$ pair transitions to a 
stop-antistop pair, which transitions to a Higgs boson, which 
transitions to a stop-antistop pair, which transitions to a $\gamma\gamma$
pair:
\begin{eqnarray}
A_4(gg\to \gamma\gamma)
&=&C_{gg\tilde t\tilde t} \tilde{G}_{\tilde t\tilde t}(\hat s) C_{H\tilde t\tilde 
t} \frac{i}{\hat s-m_H^2+im_H\Gamma_H-iC_{H\tilde t\tilde t}^2
\tilde{G}_{\tilde t\tilde t}(\hat s)}
C_{H\tilde t\tilde t} 
\tilde{G}_{\tilde t\tilde t}(\hat s) C_{\gamma\gamma\tilde t\tilde t}.
\label{A4}
\end{eqnarray}

Finally, there is a contribution that does not involve the Higgs boson:
\begin{eqnarray}
A_5(gg\to \gamma\gamma)
&=&C_{gg\tilde t\tilde t}\tilde{G}_{\tilde t\tilde t}(\hat s) 
C_{\gamma\gamma\tilde t\tilde t}.
\end{eqnarray}

We note that the amplitudes $A_4$ and $A_5$ can be combined to give a 
simpler expression:
\begin{subequations}\label{A4p}%
\begin{eqnarray}
A_4'(gg\to \gamma\gamma)&=&A_4(gg\to \gamma\gamma)+
A_5(gg\to \gamma\gamma)\\
&=&C_{gg\tilde t\tilde t}\frac{1}{\tilde{G}_{\tilde t\tilde t}^{-1}(\hat s)
-C_{H\tilde t\tilde t}^2 S_H(\hat s)} C_{\gamma\gamma\tilde t\tilde t}
\label{A4pb}\\
&=&C_{gg\tilde t\tilde t}
 \ \tilde{G}_{\tilde t\tilde t}(\hat s)
\frac{ \hat{s}-m_H^2+im_H\Gamma_H}{\hat s-m_H^2+im_H\Gamma_H
-iC_{H\tilde t\tilde t}^2\tilde{G}_{\tilde t\tilde t}(\hat s)}
C_{\gamma\gamma\tilde t\tilde t}.
\end{eqnarray}
\end{subequations}%
The form of $A_4'(gg\to \gamma\gamma)$ in Eq.~(\ref{A4pb}) is the same as
the form of $A_1(gg\to \gamma\gamma)$ in Eq.~(\ref{A1b}), but with the
r\^oles of the Higgs boson and the stop-antistop pair interchanged. 

We also note that the total amplitude
\begin{eqnarray}
A_{\rm tot}( gg \to \gamma\gamma) & = &\sum_{i=1}^5 A_i(gg\to \gamma\gamma)  
\label{A-tot}
\end{eqnarray}
can be obtained from the matrix expression 
\begin{subequations}%
\label{matrix-amps}%
\begin{eqnarray}
A_{\rm tot} (gg \to \gamma\gamma) & = &\left(
\begin{array}{cc}
C_{ggH} & C_{gg\tilde t\tilde t}
\end{array}
\right)
\left(
\begin{array}{cc}
S_H^{-1}(\hat s) & \;\;\;\; -C_{H\tilde t\tilde t}\\
-C_{H\tilde t\tilde t} & \;\;\;\;\;\; \tilde{G}_{\tilde t\tilde t}^{-1}(\hat s)
\end{array}
\right)^{-1}
\left(
\begin{array}{c}                                                        
C_{\gamma\gamma H}\label{matrix-amps-a}\\ 
C_{\gamma\gamma\tilde t\tilde t}             
\end{array}
\right)\\
&=&\left(
\begin{array}{cc}
C_{ggH} & C_{gg\tilde t\tilde t}
\end{array}
\right)
\left(
\begin{array}{cc}
\tilde{G}_{\tilde t\tilde t}^{-1}(\hat s) & \;\;\;\; C_{H\tilde t\tilde t}\\
C_{H\tilde t\tilde t} & \;\;\;\;\;\; S_H^{-1}(\hat s)
\end{array}
\right)
\left(
\begin{array}{c}                                                        
C_{\gamma\gamma H}\\ 
C_{\gamma\gamma\tilde t\tilde t}             
\end{array}
\right)\nonumber\\
&&\qquad\times \frac{1}{S_H^{-1}(\hat s)
\tilde{G}_{\tilde t\tilde t}^{-1}(\hat{s})-
C_{H\tilde t\tilde t}^2},\label{matrix-amps-b}
\end{eqnarray}
\end{subequations}%
where the matrix whose inverse is taken in Eq.~(\ref{matrix-amps-a})
is $(-i)$ times the effective Hamiltonian for the
Higgs-stop-antistop system. The expression in
Eq.~(\ref{matrix-amps-a}) is a generalization of Eq.~(7) in
Ref.~\cite{Djouadi:2016oey}.

\subsection{Coulomb-Schr\"odinger Green's function}

The stop-antistop Green's function $G_{\tilde t\tilde t}(\hat s)$
can be computed at the leading nontrivial order in $v$, by allowing the
stop and antistop to interact only through the potential of the 
leading order in $v$, which is called the static potential. One could
take the static potential from lattice data, which are well described by
a Coulomb-plus-linear potential (Cornell potential
\cite{Eichten:1974af})  with a roll-off to a flat potential above the
stop-antistop threshold. The Green's function corresponding to such a
potential could, in principle, be evaluated numerically. In this paper,
we choose instead to deal with a completely analytic Green's function,
which, we believe, illustrates the qualitative features of the
Higgs-stop-antistop system. In particular, we make use of a modified
Coulomb-Schr\"odinger Green's function, which we describe in detail
below.

We obtain the relationship between $G_{\tilde t\tilde t}(\hat s)$ and 
the Schr\"odinger Green's function as follows. Potential interactions 
are independent of the relative momentum $p_0$ of the stop quark and the 
antistop quark.
Therefore, we integrate the effective-field-theory stop and antistop
propagators over
$p_0$ to obtain
\begin{eqnarray}
&&\int\frac{dp_0}{2\pi}\frac{i}{2m_{\tilde t}(p_0+E/2)-
{\bf p}^2+im_{\tilde t}\Gamma_{\tilde t}+i\epsilon}\,
\frac{i}{-2m_{\tilde t}(p_0+E/2)-
{\bf p}^2+im_{\tilde t}\Gamma_{\tilde t}+i\epsilon}\nonumber\\
&&\qquad =\frac{i}{4m_{\tilde t}^2}
\frac{1}{E-\frac{{\bm p}^2}{2m_{\tilde t}}+i\Gamma_{\tilde t}+i\epsilon}
=\frac{-i}{4m_{\tilde t}^2}G_{\rm S}^{(0)}(E+i\Gamma_{\tilde t},{\bm p}),
\end{eqnarray}
where $\Gamma_{\tilde t}$ is the stop-quark width and 
$G_{\rm S}^{(0)}(E,{\bm p})$ is the Schr\"odinger propagator 
(Schr\"odinger Green's function in the absence of interactions).
Hence, we conclude that, in the case of a Coulomb potential, 
\begin{equation}
G_{\tilde t\tilde t}(\hat s)=\frac{-i}{4m_{\tilde t}^2}G_{\rm 
C-S}(0,0,E+i\Gamma_{\tilde t}),
\end{equation}
where $G_{\rm C-S}(0,0,E+i\Gamma_{\tilde t})$ 
is the Coulomb-Schr\"odinger Green's function 
evaluated at zero spatial separation between the initial stop and 
antistop and zero spatial separation between the final stop and       
antistop.

In the $\overline{\rm MS}$ scheme, $G_{\rm 
C-S}(0,0,E+ i  \Gamma_{\tilde t})$ is given by 
\cite{Melnikov:1998pr,Kiyo:2010jm}
\begin{subequations}\label{G-C-S}%
\begin{eqnarray}
G_{\rm C-S}(0,0,\tilde{E}) &=&
\frac{\alpha_s C_F}{4 \pi} m_{\tilde t}^2
\left[ - \frac{1}{2 \lambda} 
- \frac{1}{2} \log \left( \frac{-4 m_{\tilde t} \tilde{E}}{\mu^2} \right)
+ \frac{1}{2} +\sum_{n=1}^\infty \frac{1}{n(n/\lambda-1)} \right]
\label{G-C-Sa} 
\\
&=& \frac{\alpha_s C_F}{4 \pi} m_{\tilde t}^2
\left[ - \frac{1}{2 \lambda} 
- \frac{1}{2} \log \left( \frac{-4 m_{\tilde t}  \tilde{E}}{\mu^2} \right)
+ \frac{1}{2} - \gamma_{\rm E} - \psi (1-\lambda) \right]\label{G-C-Sb},
\end{eqnarray}
where 
\begin{equation}
\tilde{E} = E + i  \Gamma_{\tilde t},
\end{equation}
\begin{equation}
\lambda \equiv \frac{\alpha_s C_F}
{\sqrt{-4 \tilde{E}/m_{\tilde t}}},
\end{equation}
\end{subequations}%
$\mu$ is the $\overline{\rm MS}$ scale, $\gamma_{\rm E}$ is Euler's
constant, and $\psi(1-\lambda)$ is the digamma function. We take
$\mu=m_{\tilde t}$.\footnote{Another reasonable choice is
$\mu=2m_{\tilde t}$, which would shift the expressions in square
brackets in Eqs.~(\ref{G-C-Sa}) and (\ref{G-C-Sb}) by $-\log 2$. We
believe that such a shift, which is small in comparison with the term
$1/2$ in the expressions in square brackets, would have no qualitative
effect on our results.}

The first and second terms in Eq.~(\ref{G-C-Sa}) correspond to zero and
one Coulomb interaction, respectively. These are the only contributions that
are ultraviolet divergent and that, therefore, depend on the
renormalization scheme. The sum in Eq.~(\ref{G-C-Sa}) contains the
contributions involving two or more Coulomb interactions. When
$\Gamma_{\tilde t}=0$, the $n$th term in the sum contains the pole
$|\psi_{n}(0)|^2/(E_n-E)$, where $E_n$ is the energy of the $n$th bound
state  and $\psi_n(0)$ is the wave function at the origin of the
$n$th bound state. The $n$th term in the sum also contains nonpole
contributions. It is best not to separate the pole and nonpole
contributions, as either of them alone produces a spurious logarithmic
singularity at $E=0$ (threshold).

In this work, we wish to capture the essential features of the QCD
Green's function, which we expect to contain only a few bound-state
poles below threshold. Therefore, we modify $G_{\rm C-S}(0,0,E)$ by
retaining only a few terms in the sum in  Eq.~(\ref{G-C-Sa}). We expect
that, for large $m_{\tilde t}$, the lowest-lying bound states will be
given approximately by the bound states of the Coulomb potential, and so
the modified Coulomb Green's function should give a qualitative
description of the system.\footnote{Lattice measurements of the static
quark-antiquark potential \cite{Bali:2000vr}, which is identical to the
static squark-antisquark potential, suggest that the static potential is
predominantly Coulombic at short distances, as would be expected from
asymptotic freedom. A recent lattice calculation \cite{Kim:2015zqa}
indicates that the stoponium ground-state wave functions at the origin,
for $100~\hbox{GeV}\leq m_{\tilde t}\leq 750~\hbox{GeV}$, may have
values that are substantially larger than those that are obtained from
potentials that match QCD perturbation theory at short distances
\cite{Hagiwara:1990sq}. There is not, as yet, an independent
confirmation of this surprising result.}

In computing the Coulomb Green's function, we set $\alpha_s$ to a
constant by requiring that
\begin{equation}
|{\bf p}|=vm_{\tilde t}=  \alpha_s(v m_{\tilde t}) m_{\tilde t}.
\end{equation}
This equation should be approximately valid for the low-lying bound 
states. It yields $\alpha_s\approx 0.13$.

\subsection{Higgs-boson form factors \label{sec:form-factors}}

The  perturbative form factor for the coupling of the 
Higgs boson to diphotons through a stop loop is given by
\begin{equation}
F(H\to \tilde t\tilde t \to \gamma\gamma) = C_{\gamma\gamma H} + 
C_{H\tilde{t}\tilde{t}}\tilde{G}_{\tilde{t} \tilde{t}} 
C_{\gamma \gamma\tilde{t} \tilde{t}}.
\label{diphoton-form-factor}
\end{equation}
The amplitude $A_1+A_2$ is proportional to this form factor, and, hence,
this form factor is built into our formalism. There is a similar
form factor for the Higgs coupling to two gluons through a stop loop:
\begin{equation}
F(H\to \tilde t\tilde t \to gg) = C_{ggH} + 
C_{H\tilde{t}\tilde{t}}\tilde{G}_{\tilde{t} \tilde{t}} 
C_{gg\tilde{t} \tilde{t}}.
\label{gg-form-factor}
\end{equation}
The amplitude $A_1+A_3$ is proportional to this form factor, and, so, 
this form factor is also built into our formalism.
The amplitude $A_4$ is proportional to the cross term between the 
Higgs-to-diphoton form factor and the Higgs-to-digluon form factor: 
\begin{equation}
F(H\to \tilde t\tilde t \to \gamma\gamma gg) = 
C_{gg\tilde{t} \tilde{t}}
C_{H\tilde{t}\tilde{t}}^2\tilde{G}_{\tilde{t} \tilde{t}}^2 
C_{\gamma\gamma\tilde{t} \tilde{t}}.
\label{double-form-factor}
\end{equation}

The factors $\tilde{G}_{\tilde t\tilde t}$ in the second terms of the
form factors in Eqs.~(\ref{diphoton-form-factor}) and
(\ref{gg-form-factor}) and in the form-factor contribution in
Eq.~(\ref{double-form-factor}) give enhancements of the total
amplitude when $\hat{s}$ is near a threshold peak in $\tilde{G}_{\tilde
t\tilde t}$. Such enhancements are already present in the perturbative
calculations that make use of  the first and second terms of the Coulomb
Green's function [Eq.~(\ref{G-C-Sa})], and they become stronger when one
takes into account the additional effects that are associated with the
stop-antistop bound states~\cite{Melnikov:1994jb}. However, as we will
illustrate in detail in Sec.~\ref{sec:breit-wigner}, when the Higgs mass
is close to stop-antistop threshold, the effect of Higgs-stop-antistop
mixing is to displace the physical mass peaks away from threshold.
Consequently, when the full effects of Higgs-stop-antistop mixing are
taken into account,  this enhancement effect is not operative.

\section{Case of a single Breit-Wigner resonance in the 
stop-antistop Green's function \label{sec:breit-wigner}}

We now discuss the situation in which we model $\tilde{G}_{\tilde
t\tilde t}$ with the simplified form of a Breit-Wigner
resonance.\footnote{This simplified model has also been considered in
Ref.~\cite{Drees:1993uw}. In Ref.~\cite{Baumgart:2012pj}, this model
was used to investigate the effects of mixing of a light pseudoscalar
particle with $\eta_b(n)$ states in the context of decays of the
pseudoscalar particle to the $\eta_b(n)$ states.}  This is the only
modification to the effective theory that we make. In particular, we use
the formulas for the short-distance coefficients that are given above.

\subsection{Structure of the Breit-Wigner amplitude}

We consider the situation in which $\tilde{G}_{\tilde t\tilde t}$ is given by 
\begin{subequations}%
\begin{equation}
\tilde{G}_{\tilde t\tilde t}=N_{\tilde t\tilde t}^{2}
S_{\tilde t \tilde t}(\hat{s}),
\end{equation}
where
\begin{equation}
S_{\tilde t\tilde t}(\hat{s})=
\frac{i}
{\hat{s}-m_{\tilde t\tilde t}^2+im_{\tilde t\tilde t}
\Gamma_{\tilde t\tilde t}},
\label{Breit-Wigner}
\end{equation}
\end{subequations}%
$m_{\tilde t\tilde t}$ is the bound-state mass, $\Gamma_{\tilde 
t\tilde t}$ is the bound-state width,
\begin{equation}
N_{\tilde t\tilde t}^{2}=
|\psi(0)|^2/m_{\tilde t},
\label{Ntt}
\end{equation}
and $\psi(0)$ is the bound-state wave function at the origin. 
The bound-state width is given by 
\begin{subequations}%
\begin{equation}
\Gamma_{\tilde t\tilde t}=2\Gamma_{\tilde t}+\Gamma_{\tilde t\tilde t gg}. 
\end{equation}
Here, $\Gamma_{\tilde t\tilde tgg}$, is the width of the bound
state to two gluons :
\begin{equation}
\Gamma_{\tilde t\tilde tgg}=
\frac{1}{(2m_{\tilde t})^2} 
2{\rm Im}\, T_{\tilde t\tilde t\to gg\to \tilde t\tilde t}\, |\psi(0)|^2
=\frac{4\pi\alpha_s^{2}(m_{\tilde t})|\psi(0)|^2}{3m_{\tilde t}^2}.
\end{equation}
\end{subequations}
For the Coulomb ground state,  
$|\psi(0)|^2=8\alpha_s^3(m_{\tilde t}v)m_{\tilde t}^3/(27\pi)$.

Now, we can write Eq.~(\ref{matrix-amps-a}) as
\begin{eqnarray}
A_{\rm tot} (gg \to \gamma\gamma)
&=&\left(
\begin{array}{cc}
C_{ggH} & \hat{C}_{gg\tilde t\tilde t}
\end{array}
\right)
\left(
\begin{array}{cc}
S_{H}^{-1}(\hat s) & \;\;\;\; -\hat{C}_{H\tilde t\tilde t}\\
-\hat{C}_{H\tilde t\tilde t} & \;\;\;\;\;\; S_{\tilde t\tilde t}^{-1}(\hat s)
\end{array}
\right)^{-1}
\left(
\begin{array}{c}                                                        
C_{\gamma\gamma H}\\ 
\hat{C}_{\gamma\gamma\tilde t\tilde t}             
\end{array}
\right),
\label{matrix-amps-one-pole}
\end{eqnarray}
where $\hat{C}_{gg\tilde t\tilde t}=N_{\tilde t\tilde t}C_{gg\tilde t\tilde 
t}$, $\hat{C}_{\gamma\gamma\tilde t\tilde t}=
N_{\tilde t\tilde t}C_{\gamma\gamma\tilde t\tilde t}$, and
$\hat{C}_{H\tilde t\tilde t}=N_{\tilde t\tilde t}
C_{H\tilde t\tilde t}$.
We can diagonalize the matrix in Eq.~(\ref{matrix-amps-one-pole}) by 
making use of a similarity transformation:
\begin{equation}
-i\left(
\begin{array}{cc}
\hat{s}^2-m_{+}^2+im_+\Gamma_+ & \;\;\;\; 0\\
0 & \;\;\;\;\;\; \hat{s}^2-m_-^2+im_-\Gamma_-
\end{array}
\right)
=
S(\theta)^{-1}\left(
\begin{array}{cc}
S_{H}^{-1}(\hat s) & \;\;\;\; \hat{C}_{H\tilde t\tilde t}\\
\hat{C}_{H\tilde t\tilde t} & \;\;\;\;\;\; S_{\tilde t\tilde t}^{-1}(\hat s)
\end{array}
\right)S(\theta),
\label{diagonal-matrix}
\end{equation}
which implies that
\begin{subequations}%
\begin{eqnarray}
A_{\rm tot} (gg \to \gamma\gamma)
&=&\left(
\begin{array}{cc}
C_{gg+} & C_{gg-}
\end{array}
\right)
\left(
\begin{array}{cc}
\frac{i}{\hat{s}^2-m_+^2+im_+\Gamma_+}& \;\;\;\;0\\
0 & \;\;\;\;\;\;\frac{i}{\hat{s}^2-m_-^2+im_-\Gamma_-}
\end{array}
\right)
\left(
\begin{array}{c}                                                        
C_{\gamma\gamma +}\\ 
C_{\gamma\gamma -}             
\end{array}
\right),
\label{diag-amp}
\end{eqnarray}
where 
\begin{equation}
\left(
\begin{array}{c}                                                        
C_{\gamma\gamma +}\\
C_{\gamma\gamma -}             
\end{array}
\right)
=S^{-1}(\theta)
\left(
\begin{array}{c}                                                        
C_{\gamma\gamma H}\\
\hat{C}_{\gamma\gamma\tilde t\tilde t}             
\end{array}
\right),
\end{equation} 
and
\begin{equation}
\left(\begin{array}{cc}
C_{gg+} & C_{gg-}
\end{array}
\right)
=\left(\begin{array}{cc}
C_{ggH} & \hat{C}_{gg\tilde t\tilde t}
\end{array}
\right)
S(\theta).
\end{equation}
\end{subequations}%

The masses and widths are given by
\begin{eqnarray}
m_\pm^2-im_\pm\Gamma_\pm&\equiv&
\tfrac{1}{2}(m_H^2-im_H\Gamma_H+m_{\tilde t\tilde t}^2-im_{\tilde t\tilde t}
\Gamma_{\tilde t\tilde t})\nonumber\\
&&\pm \tfrac{1}{2}\sqrt{(m_H^2-im_H\Gamma_H-m_{\tilde t\tilde t}^2+
im_{\tilde t\tilde t}\Gamma_{\tilde t\tilde t})^2
+4|\hat{C}_{H\tilde t\tilde t}|^2}.
\end{eqnarray}
(Recall that, in our definition, $\hat{C}_{H\tilde t\tilde t}$ is purely
imaginary.) Note that these values of $m_\pm-im_\pm \Gamma_\pm$ 
correspond precisely to the values of $\hat{s}$ at which the 
denominators in $A_1$\ldots $A_4$ [Eqs.~(\ref{A1})--(\ref{A4}) and 
Eq.~(\ref{matrix-amps-b})] vanish.
 
Approximate expressions for the masses and widths are
\begin{subequations}%
\label{masses-widths}
\begin{equation}
m_\pm^2\approx
\tfrac{1}{2}(m_H^2+m_{\tilde t\tilde t}^2)
\pm \tfrac{1}{2} \Delta,
\end{equation}
\begin{equation}
m_\pm\Gamma_\pm \approx
\tfrac{1}{2}\biggl[m_H\Gamma_H
+m_{\tilde t\tilde t}\Gamma_{\tilde t\tilde t}
\pm (m_H\Gamma_H-m_{\tilde t\tilde t}\Gamma_{\tilde t\tilde t})
\frac{m_H^2-m_{\tilde t\tilde t}^2}{\Delta}\biggr],
\end{equation}
where
\begin{equation}
\Delta\equiv
\sqrt{(m_H^2-m_{\tilde t\tilde t}^2)^2
+ 4 |\hat{C}_{H\tilde t\tilde t}|^2}.
\end{equation}
\end{subequations}
In the approximate forms in Eq.~(\ref{masses-widths}), we have neglected
terms of higher order in $(m_H\Gamma_H-m_{\tilde t\tilde
t}\Gamma_{\tilde t\tilde t})^2/\Delta^2$, which is less than $2\%$ for
the values of $m_H\Gamma_H$, $m_{\tilde t\tilde t}\Gamma_{\tilde t\tilde
t}$, and $\hat{C}_{H\tilde t\tilde t}^2$ that we use in our
cross-section calculations in Sec.~\ref{sec:breit-wigner-qualitative}.

The matrix $S(\theta)$ is given by
\begin{equation}
S(\theta)=
\left(
\begin{array}{cc}
\cos\theta& \;\;\;\; -\sin\theta \\
\sin\theta & \;\;\;\;\;\; \cos\theta
\end{array}
\right),
\end{equation}
where
\begin{subequations}%
\begin{eqnarray}
\tan\theta&=&\frac{2|\hat{C}_{H\tilde t\tilde t}|}
{\sqrt{(m_H^2-im_H\Gamma_H-m_{\tilde t\tilde t}^2+
im_{\tilde t\tilde t}\Gamma_{\tilde t\tilde t})^2
+4|\hat{C}_{H\tilde t\tilde t}|^2}
-(m_H^2-im_H\Gamma_H-m_{\tilde t\tilde t}^2+        
im_{\tilde t\tilde t}\Gamma_{\tilde t\tilde t})}\nonumber\\
&&\approx
\frac{2|\hat{C}_{H\tilde t\tilde t}|}
{\Delta -(m_H^2-m_{\tilde t\tilde t}^2)+i\delta
},
\label{similarity-angle}
\end{eqnarray}
with
\begin{equation}
\delta=(m_H\Gamma_H -m_{\tilde t\tilde t}\Gamma_{\tilde t\tilde t})
\biggl(1-\frac{m_H^2-m_{\tilde t\tilde t}^2}{\Delta}\biggr).
\end{equation}
\end{subequations}%
In the approximation in Eq.~(\ref{similarity-angle}), we have again
neglected terms of higher order in $(m_H\Gamma_H-m_{\tilde t\tilde
t}\Gamma_{\tilde t\tilde t})^2/\Delta^2$.

Note that $m_+^2$ and $m_-^2$ are always centered at the average of
$m_H^2$ and $m_{\tilde t\tilde t}^2$ and separated from each other by a
nonzero amount, namely, $\Delta$. When $2|\hat{C}_{H\tilde t\tilde t}|$
is small in comparison with $|m_H^2-m_{\tilde t\tilde t}^2|$, $\theta$
approaches 0 or $\pi/2$, and the masses and widths approach their
original values. On the other hand, when $|m_H^2-m_{\tilde t\tilde
t}^{2}|$ is negligible in comparison with $2|\hat{C}_{H\tilde t\tilde t}|$,
mixing is maximal, and $\theta$ is very close to
$\pi/4$.\footnote{For the values of the theory parameters that are
given in Sec.~\ref{sec:breit-wigner-qualitative}, the real part of
$\theta$ is within $1\%$ of $\pi/4$, while the imaginary part of
$\theta$ is less than $7\%$.} In this case of maximal mixing, $\Delta$
approaches its minimum value, $2|\hat{C}_{H\tilde t\tilde t}|$, and
\begin{equation}
m_\pm^2\approx \tfrac{1}{2}(m_H^2+m_{\tilde t\tilde t}^2)\pm 
|\hat{C}_{H\tilde t\tilde t}|.
\end{equation}
In the case of maximal mixing, the widths are given by
\begin{equation}
m_{\pm} \Gamma_\pm\approx \tfrac{1}{2}(m_H\Gamma_H
+m_{\tilde t\tilde t}\Gamma_{\tilde t\tilde t}).
\end{equation}
That is, the widths $\Gamma_+$ and $\Gamma_-$ become approximately the
average of $\Gamma_H$ and $\Gamma_{\tilde t\tilde t}$.

It is illuminating to compute the physical mass separations and widths
for the values of the theory parameters that we use in our
cross-section calculations in Sec.~\ref{sec:breit-wigner-qualitative}.
Even in the case of maximal mixing, for which the separation between
$m_+$ and $m_-$ is minimal, that separation is still substantial: about
$4.7$~GeV for $\kappa=1$ and about $37.4$~GeV for $\kappa=8$. These
separations are much greater than the widths of the Breit-Wigner
resonances, which are $\Gamma_H\approx 1.2$~GeV, $\Gamma_{\tilde t\tilde
t}\approx 0.3$~MeV for $\Gamma_{\tilde t}=0.1$~MeV, and $\Gamma_{\tilde
t\tilde t}\approx 0.2$~GeV for $\Gamma_{\tilde t}=0.1$~GeV.
Consequently, the mass separations are also much greater than
$\Gamma_\pm$. That is, the physical resonances are well separated for
any value of $m_H$ relative to $m_{\tilde t\tilde t}$.

In Sec.~\ref{sec:form-factors}, we discussed the threshold enhancements
that are present in the form factors for the Higgs couplings to $gg$ and
$\gamma\gamma$. We now see that those enhancements are rendered
inoperative because, when $m_H$ approaches the stop-antistop threshold,
the physical masses are displaced from threshold by a sizable amount.
In the next section, we will provide a detailed numerical analysis
of this effect.

\subsection{Qualitative features of the Breit-Wigner cross 
section\label{sec:breit-wigner-qualitative}}

In this section, we present numerical results for the short-distance
coefficients and also for cross sections at the LHC at a CM energy
$\sqrt{s}$ of $13$~TeV. We note that the $gg$ contribution to the
total cross section is given by the expression
\begin{equation}
\sigma_{\rm tot}=\frac{1}{128\pi s}
\int_{\sqrt{\hat{s}_{\rm min}}}^{\sqrt{\hat{s}_{\rm max}}} 
\frac{d\sqrt{\hat{s}}}{\sqrt{\hat{s}}} \int_{\hat{s}/s}^1 
\frac{dx}{x}f_g(\tfrac{\hat{s}}{xs}) f_g(x) |A_{\rm tot}(\hat{s})|^2,
\label{sigmatot}
\end{equation}
where $f_g$ is the gluon distribution and $\sqrt{\hat{s}_{\rm min}}$
and $\sqrt{\hat{s}_{\rm max}}$ are the lower and upper limits,
respectively, of the range in $\sqrt{\hat{s}}$ that includes  all
relevant contributions to the cross section.  In the computations of
$\sigma_{\rm tot}$ in the remainder of this paper, we take
$\sqrt{\hat{s}_{\rm min}}=600$~GeV and $\sqrt{\hat{s}_{\rm
max}}=900$~GeV.

\begin{figure}
\epsfig{file=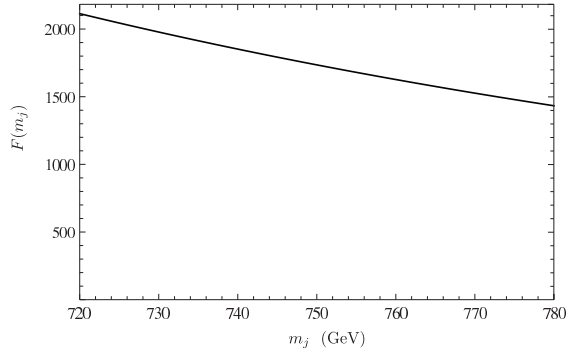,width=0.45\columnwidth} 
\caption{\label{fig:gluon-flux}%
The gluon flux factor $F(m_j)$. }
\end{figure}

As we have already mentioned, the Breit-Wigner resonances are always
well separated in comparison to their widths. Therefore, we can
approximate the cross section in Eq.~(\ref{sigmatot}) as a sum of the
individual contributions of the two resonances. If we also neglect the
dependences of the gluon distributions on $\hat{s}$ over the width of
the resonance, then we obtain the narrow-resonance approximation
\begin{subequations}%
\begin{eqnarray}
\sigma_{\rm tot}&=&
\frac{1}{128\pi s}\sum_{j=\pm} 
\int_0^\infty \frac{d\sqrt{\hat{s}}}{m_j}\,
\biggl|\frac{i}{\hat{s}-m_j^2+im_j\Gamma_j}\biggr|^2
|C_{ggj}C_{\gamma\gamma j}|^2 F(m_j) +O(\Gamma_j/m_j) \nonumber\\
&&=\frac{1}{256s}\sum_{j=\pm} 
\frac{|C_{ggj}C_{\gamma\gamma j}|^2 }{m_j^3 \Gamma_j}F(m_j)
+O(\Gamma_j/m_j),
\label{sigma-tot-approx}
\end{eqnarray}
where $F(m_j)$ is the gluon flux factor:
\begin{equation}
F(m_j)=\int_{m_j^2/s}^1 \frac{dx}{x}f_g(\tfrac{m_j^2}{xs}) f_g(x).
\end{equation}
\end{subequations}
As can be seen from Fig.~\ref{fig:gluon-flux}, $F(m_j)$ is a slowly
varying function of $m_j$ over the range of interest.\footnote{In the
calculations of cross sections in this paper, we make use of the CTEQ6M
parton distribution functions~\cite{Pumplin:2002vw}.} It has only a
small effect on the shape of the cross section as a function of $m_H$,
which is determined mainly by the dependences of the short-distance
coefficients $C_{ggj}$ and $C_{\gamma\gamma j}$ and the resonance widths
$\Gamma_j$ on $m_H$.

Note that the cross section in the narrow-resonance approximation varies
inversely as the width $\Gamma_j$. In general, the cross-section
contribution of a Breit-Wigner resonance is proportional to the square
of the absolute value of the maximum height of the amplitude times the
width of the resonance.  In the Breit-Wigner model of this section,
the inverse of the diagonal matrix in Eq.~(\ref{diagonal-matrix}) has
elements whose maximum absolute values are equal to $1/(m_j\Gamma_j)$.
Therefore, the effect of this matrix factor on the cross section can be
characterized completely in terms of the widths of the physical
resonances. We use this characterization for the remainder of this
section. As we will explain in Sec.~\ref{sec:differences}, in case of
the Coulomb-Schr\"odinger model, the corresponding diagonal matrix is
more complicated, and its effect on the cross section cannot be
characterized completely in terms of the widths of the physical
resonances.

We discuss below the behavior of the cross section for four 
cases: $\kappa=1$, $\Gamma_{\tilde t}=0.1$~MeV; 
$\kappa=1$, $\Gamma_{\tilde t}=0.1$~GeV; 
$\kappa=8$, $\Gamma_{\tilde t}=0.1$~MeV; and
$\kappa=8$, $\Gamma_{\tilde t}=0.1$~GeV.
The $\kappa$ value of $8$ is near the upper limit of the values of $\kappa$
that are allowed by unitarity constraints \cite{Allanach:2015blv}.
As we have mentioned, in order to make contact with previous numerical
work, we take the stop mass to be $m_{\tilde t} = 375$~GeV. We also use
the input values $m_b(2m_{\tilde t})=2.46$~GeV, $m_t(2m_{\tilde
t})=149.95$~GeV, where $m_t(2 m_{\tilde t})$ and $m_b(2 m_{\tilde t})$
are the $\overline{\rm MS}$ running masses at the scale $2 m_{\tilde t}$. We
estimate the Higgs width by computing its widths to top- and
bottom-quark pairs through order
$\alpha_s$~\cite{Braaten:1980yq,Drees:1990dq}\footnote{We have converted the
result in Ref.~\cite{Drees:1990dq}, which is expressed in terms of quark
pole mass, to an expression in terms of the quark
modified-minimal-subtraction ($\overline{\rm MS}$) mass by adding
$2+(3/2)\log[4/(1-\beta^2)]$ to $\Delta_H$ in Eq.~(2.26) of
Ref.~\cite{Drees:1990dq}.} and its width to a $\tau$-lepton pair at the
Born level, using the Higgs couplings to top-quark, bottom-quark, and
$\tau$-lepton pairs that occur in the decoupling limit of large Higgs
mass \cite{Branco:2011iw}. We will make use of an intermediate value
of $\tan\beta$, setting $\tan\beta=\sqrt{m_t(2 m_{\tilde t})/m_b(2
m_{\tilde t})}$. At this value of $\tan\beta$, the Higgs decay width
into third generation fermions is minimized, and our estimate is
$\Gamma_H  \simeq 1.2$~GeV.

\subsubsection{$\kappa=1$, $\Gamma_{\tilde 
t}=0.1$~MeV\label{sec:breit-wigner-1n}}

We consider first the case of weak Higgs-stop-antistop coupling, 
$\kappa=1$, and small stop width, $\Gamma_{\tilde t}=0.1$~MeV.

\begin{figure}
\epsfig{file=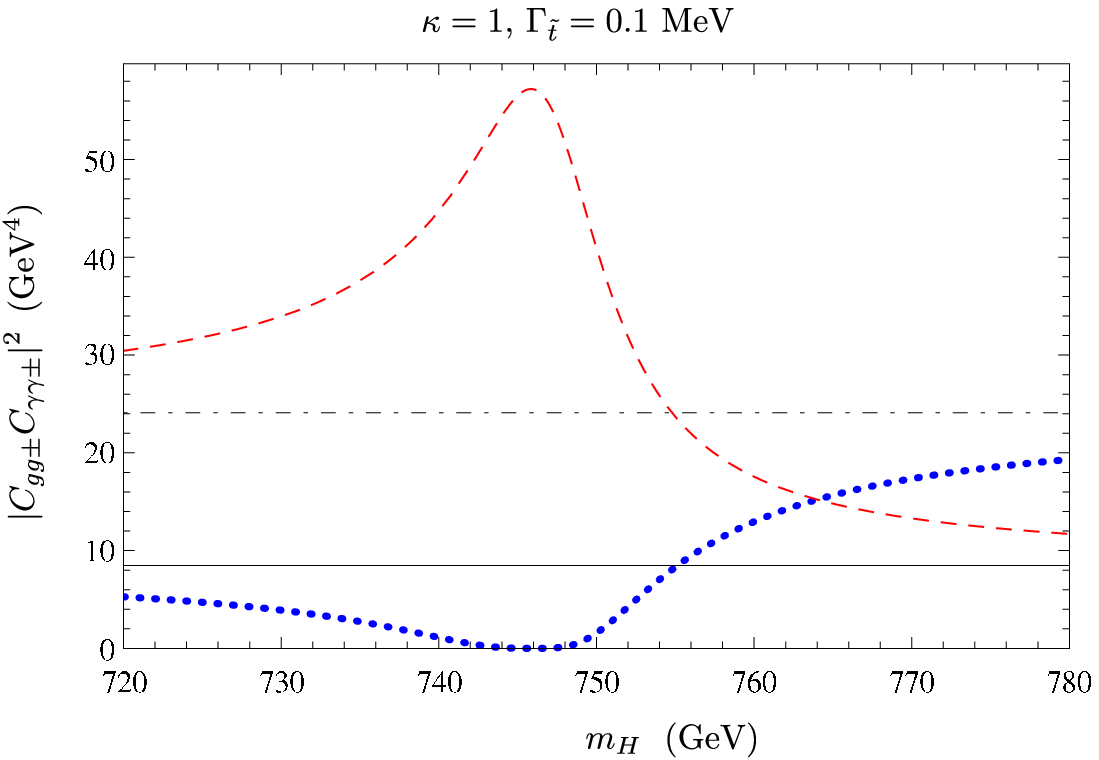,width=0.45\columnwidth} 
\epsfig{file=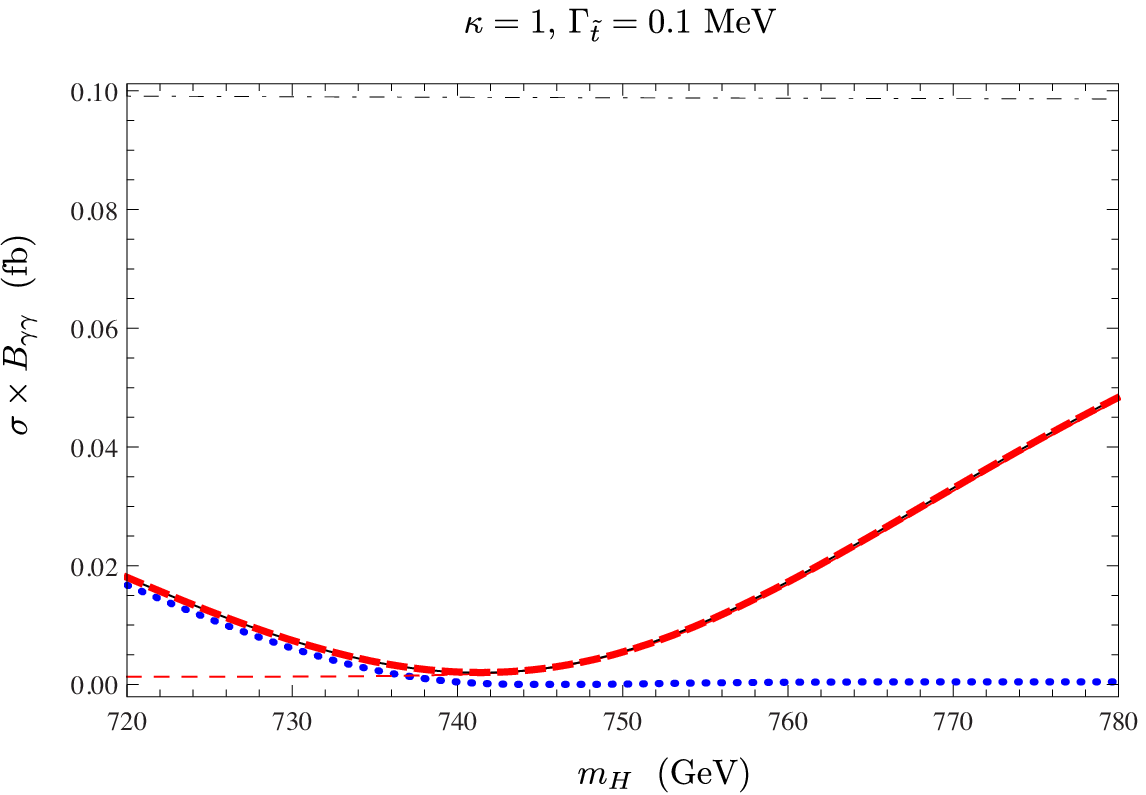,width=0.45\columnwidth}
\caption{\label{fig:BR-1n}%
Numerical results for the case in which the stop-antistop Green's
function is given by a Breit-Wigner resonance for $\kappa=1$ and
$\Gamma_{\tilde t}=0.1$~MeV. Quantities are presented as functions of
$m_H$. The left panel shows the products of absolute squares of
short-distance coefficients: $|C_{gg+}C_{\gamma\gamma +}|^2$ (dotted,
blue line), $|C_{gg-}C_{\gamma\gamma -}|^2$ (narrow, dashed, orange
line), $|C_{gg\tilde t\tilde t}C_{\gamma\gamma\tilde t\tilde t}|^2$
(solid, black line), and $|C_{ggH}C_{\gamma\gamma H}|^2$ (dash-dotted, black
line). The right panel shows cross sections times branching ratios 
into $\gamma\gamma$: contribution in the narrow-resonance approximation
of the larger-mass eigenstate (dotted, blue line), contribution in the
narrow-resonance approximation of the smaller-mass eigenstate (narrow,
dashed, orange line), sum of the contributions in the narrow-resonance
approximation (thick, dashed, red line), exact cross section times 
branching ratio into $\gamma\gamma$ (thin, black line), and exact cross
section times branching ratio into $\gamma\gamma$ in the absence of
mixing (dash-dotted, black line).}
\end{figure}

Let us examine the effect of mixing on the short-distance 
coefficients. 
In the left panel of Fig.~\ref{fig:BR-1n}, we show $|C_{\gamma\gamma
\pm}C_{gg\pm}|^2$ as functions of $m_H$. We also show
$|C_{\gamma\gamma \tilde t\tilde t}C_{gg\tilde t\tilde t}|^2$ and
$|C_{\gamma\gamma H}C_{ggH}|^2$, so that one can judge the importance of
the mixing effects.
At small values of $m_H$, the lower-mass eigenstate corresponds to the
Higgs boson, and the higher-mass eigenstate corresponds to the
stop-antistop bound state. This correspondence is reversed for large
values of $m_H$.  That is, at
both large and small values of $m_H$, the upper line corresponds to the
Higgs coefficients, and the lower line corresponds to the stop-antistop
coefficients. Maximal mixing occurs when $m_H$ is equal to the
stop-antistop bound-state mass $m_{\tilde t\tilde t}$, which is about
$747$~GeV. For large values of $|m_H-m_{\tilde t\tilde t}|$, the
mixing angle decreases approximately as $1/|m_H-m_{\tilde t\tilde t}|$.
However, we remind the reader that the nonrelativistic approximation
that we use in our calculations is valid only when $|m_H-m_{\tilde
t\tilde t}|\ll m_{\tilde t\tilde t}$. The structure that appears in
the quantities $|C_{gg\pm}C_{\gamma\gamma \pm}|^2$ is a consequence of
the fact that the Higgs coefficients contain real and imaginary parts
that are comparable in magnitude. The mixing then produces a complicated
pattern of interference. We emphasize that the peak that appears in the
upper line is not produced by a resonance or by a threshold enhancement
from the Higgs to $gg$ or $\gamma\gamma$ form factors. Rather, it is
entirely a consequence of interference effects in the short-distance
coefficients.

Now, let us consider the behavior of the cross section as a function of
$m_H$. In the right panel of Fig.~\ref{fig:BR-1n}, we show the
contributions of the larger-mass eigenstate and the smaller-mass
eigenstate to the cross section times the branching ratio into
$\gamma\gamma$ in the narrow-resonance approximation, the sum of those
contributions, and the exact cross section times the branching ratio into 
$\gamma\gamma$. We also show the exact cross section in the absence of 
mixing so that one can judge the importance of the mixing effects.

We can understand the qualitative features of the cross section times the 
branching ratio into $\gamma\gamma$ from the formula for the
narrow-resonance approximation to the cross section in
Eq.~(\ref{sigma-tot-approx}). As can be seen from Fig.~\ref{fig:BR-1n},
the factors $1/\Gamma_\pm$ in Eq.~(\ref{sigma-tot-approx}) have a strong
effect on the shape of the cross section. When $m_H$ is small (large),
the mixing is minimal, and $\Gamma_+$ ($\Gamma_-$) is equal to
$\Gamma_{\tilde t\tilde t}\approx 3$~MeV. When $m_H$ approaches
$m_{\tilde t \tilde t}$, the mixing becomes maximal, and $\Gamma_\pm
\approx (\Gamma_H+\Gamma_{\tilde t\tilde t})/2\approx 0.6$~GeV. The
dependence of the cross section on $1/\Gamma_\pm$ completely overwhelms
the dependence of the cross section on the short-distance coefficients,
resulting in the shape for the total cross section that is shown in the
thick, dashed, red line in the right panel. The cross section times
the branching ratio is suppressed at values of $m_H$ that are close to
$m_{\tilde t \tilde t}$, owing to the increase in the width of the
narrowest resonance from the stoponium width to the average of the
stoponium and Higgs widths. Even at $m_H=720$~GeV and $m_H=780$~GeV,
where the mixing angles are fairly small, the effect of mixing on the
width of the narrowest resonance leads to a considerable suppression of
the values of the cross section times the branching ratio relative to the
values in the absence of mixing. We note
that the narrow-resonance approximation gives a result for the cross
section that is very close to the exact result. (The line for the exact
result is almost completely obscured by the line for the
narrow-resonance-approximation result.)

\subsubsection{$\kappa=1$, $\Gamma_{\tilde
t}=0.1$~GeV\label{sec:numerical-1n}}

Next, we consider the case of weak Higgs-stop-antistop coupling,
$\kappa=1$, and large stop width, $\Gamma_{\tilde t}=0.1$~GeV. 

\begin{figure}
\epsfig{file=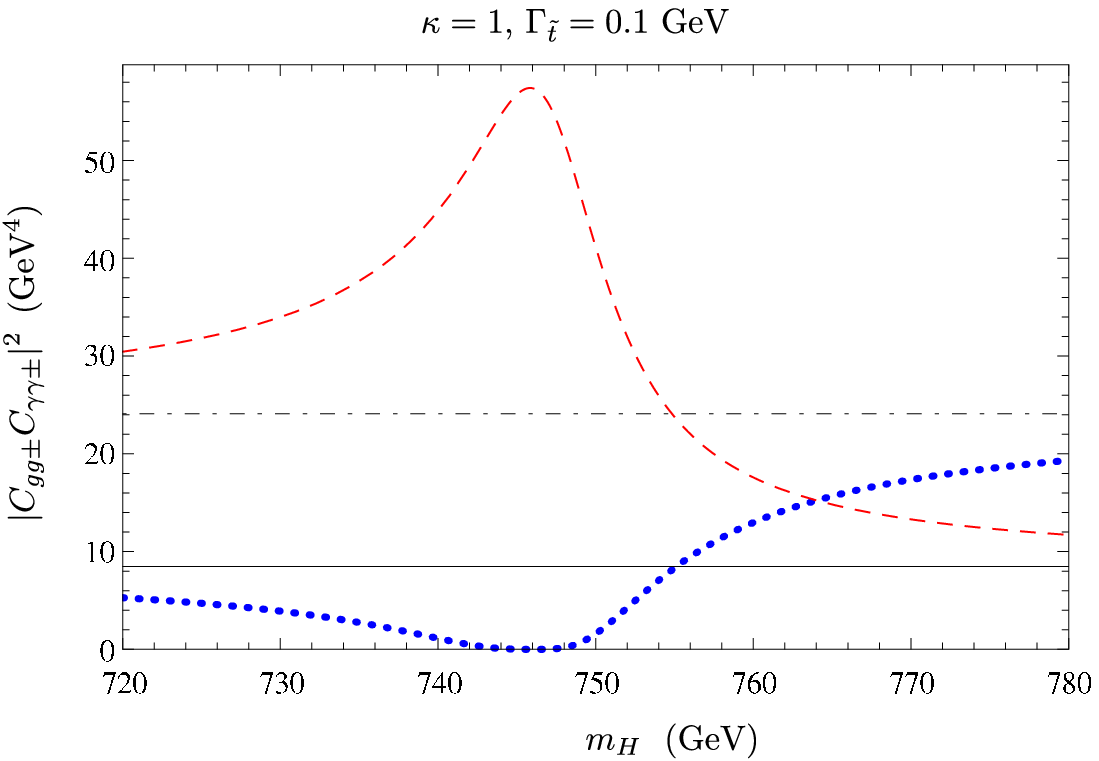,width=0.45\columnwidth} 
\epsfig{file=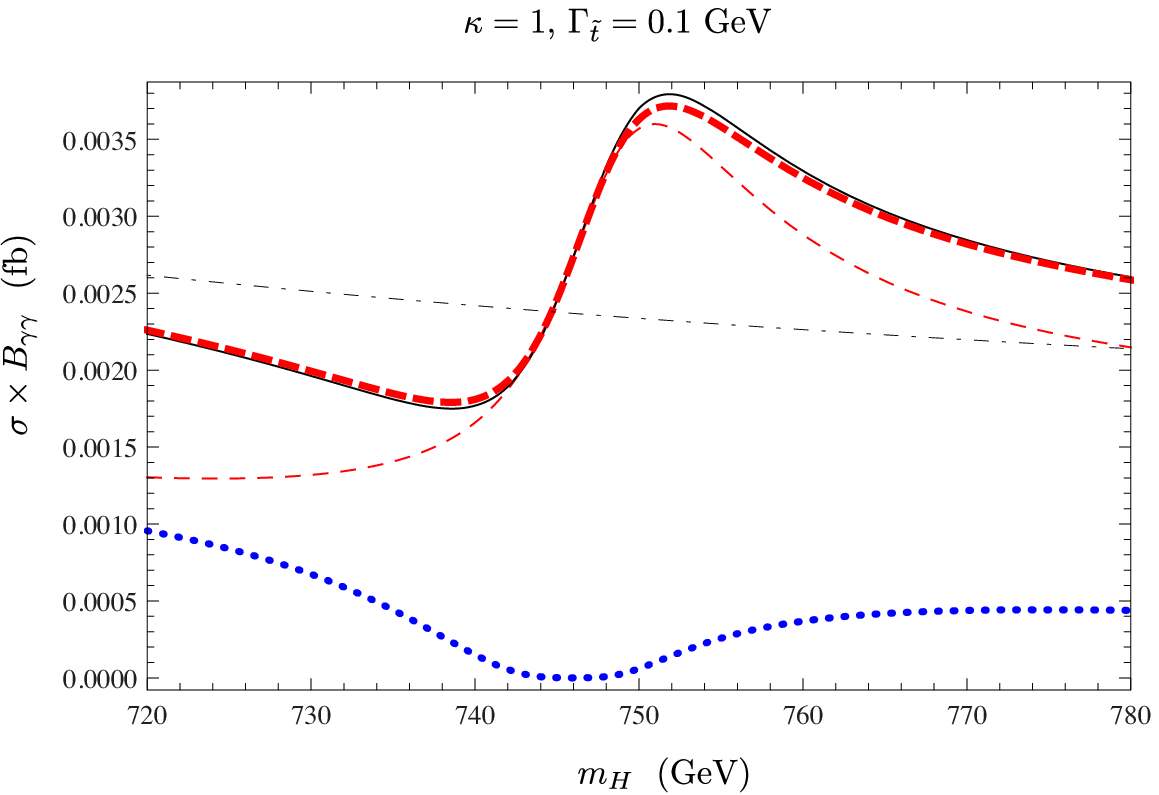,width=0.45\columnwidth}
\caption{\label{fig:BR-1w}%
Numerical results for the case in which the stop-antistop Green's
function is given by a Breit-Wigner resonance for $\kappa=1$ and
$\Gamma_{\tilde t}=0.1$~GeV. Quantities are presented as functions of
$m_H$. The left panel shows the products of absolute squares of
short-distance coefficients: $|C_{gg+}C_{\gamma\gamma +}|^2$ (dotted,
blue line), $|C_{gg-}C_{\gamma\gamma -}|^2$ (narrow, dashed, orange
line), $|C_{gg\tilde t\tilde t}C_{\gamma\gamma\tilde t\tilde t}|^2$
(solid, black line), and $|C_{ggH}C_{\gamma\gamma H}|^2$ (dash-dotted, black
line). The right panel shows cross sections times branching ratios 
into $\gamma\gamma$: contribution in the narrow-resonance approximation
of the larger-mass eigenstate (dotted, blue line), contribution in the
narrow-resonance approximation of the smaller-mass eigenstate (narrow,
dashed, orange line), sum of the contributions in the narrow-resonance
approximation (thick, dashed, red line), exact cross section times 
branching ratio into $\gamma\gamma$ (thin, black line), and exact cross
section times branching ratio into $\gamma\gamma$ in the absence of
mixing (dash-dotted, black line).}
\end{figure}

In Fig.~\ref{fig:BR-1w}, we display the values of $|C_{\gamma\gamma
\pm}C_{gg\pm}|^2$ as functions of $m_H$. Again, we also show
$|C_{\gamma\gamma \tilde t\tilde t}C_{gg\tilde t\tilde t}|^2$ and
$|C_{\gamma\gamma H}C_{ggH}|^2$, so that one can judge the importance of
the mixing effects.  As can be seen by comparing the left panel of
Fig.~\ref{fig:BR-1w} with the left panel of Fig.~\ref{fig:BR-1n}, the
effect of mixing on the short-distance coefficients in the case of large
stop width is essentially the same as in the case of small stop width.
However, the stop-antistop width is now $\Gamma_{\tilde t\tilde t}
\approx 0.2$~GeV, which is not far from the Higgs width $\Gamma_H\approx
1.2$~GeV. It follows that the width of the narrowest resonance goes from
about $0.2$~GeV for minimal mixing to about $0.7$~GeV for maximal
mixing, which is a much smaller range than in the case of a small stop
width. Consequently, as can be seen from the right panel of
Fig.~\ref{fig:BR-1w}, the effect of mixing on the resonance widths has a
much less dramatic effect on the shape of the cross section times
the branching ratio than in the small-stop-width case. The shape of the
cross section times the branching ratio in the thick, dashed, red line now
exhibits a peak that corresponds to the peak in the sum of the
magnitudes of the products of short-distance coefficients. Comparing
with the situation for small stop width, we see that the cross section
times the branching ratio away from threshold is significantly smaller,
owing to that fact that the stoponium width is now much larger. For the
same reason, the cross section times the branching ratio away from threshold
more closely approaches the cross section in the absence of
mixing. Again, we note that the result from the narrow-resonance
approximation for the cross section times the branching ratio into
$\gamma\gamma$ agrees well with the exact result.

\subsubsection{$\kappa=8$, $\Gamma_{\tilde t}=0.1$~MeV}

Next, we consider the case of strong Higgs-stop-antistop coupling, 
$\kappa=8$, and small stop width, $\Gamma_{\tilde t}=0.1$~MeV.

\begin{figure}
\epsfig{file=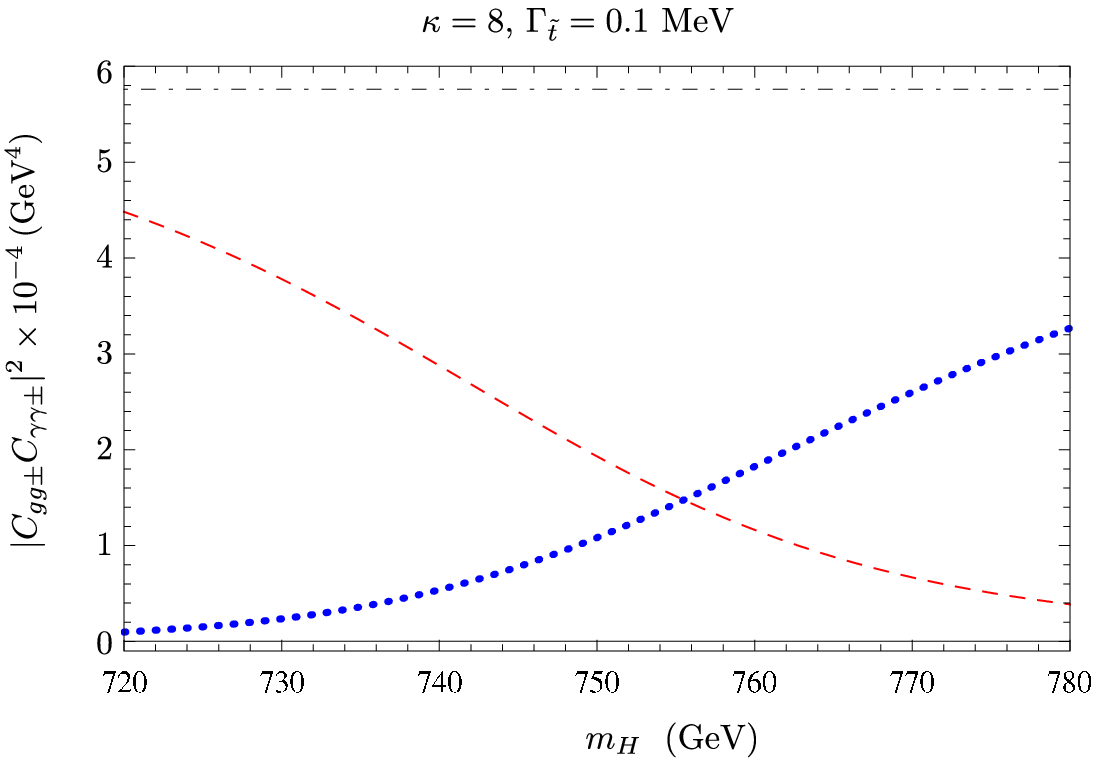,width=0.45\columnwidth} 
\epsfig{file=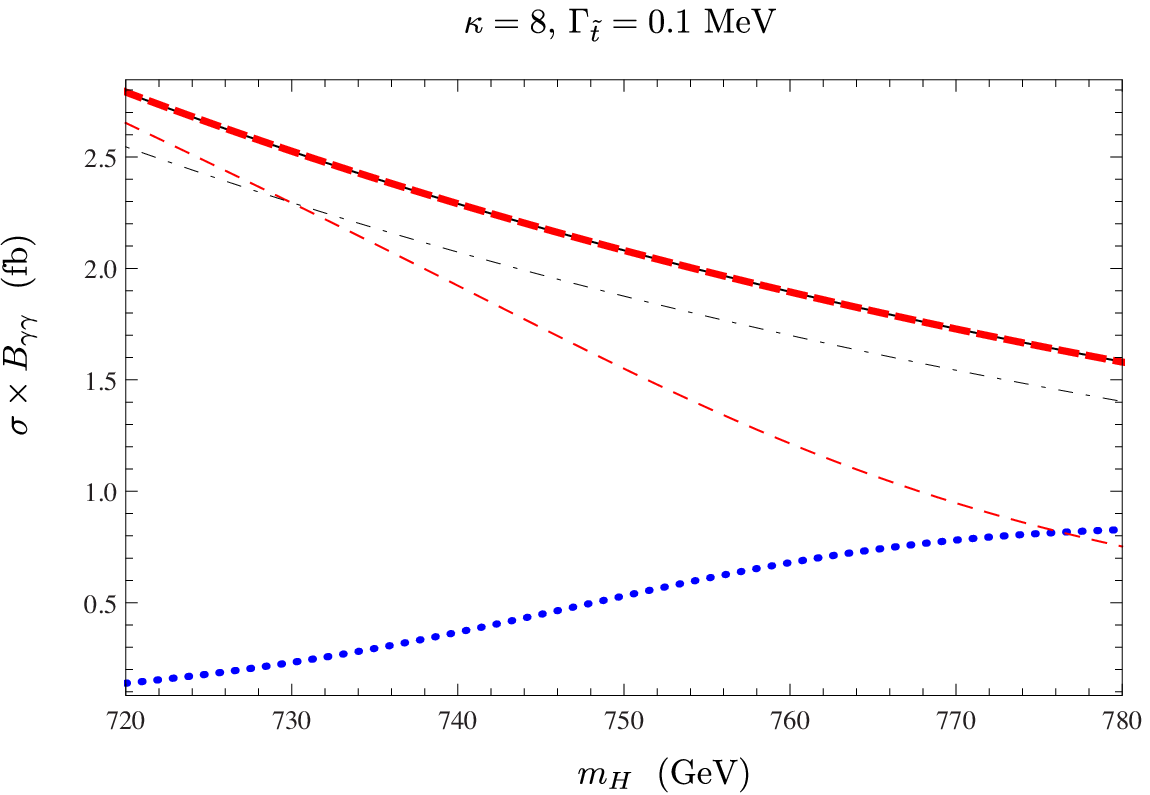,width=0.45\columnwidth}
\caption{\label{fig:BR-8n}%
Numerical results for the case in which the stop-antistop Green's
function is given by a Breit-Wigner resonance for $\kappa=8$ and
$\Gamma_{\tilde t}=0.1$~MeV. Quantities are presented as functions of
$m_H$. The left panel shows the products of absolute squares of
short-distance coefficients: $|C_{gg+}C_{\gamma\gamma +}|^2$ (dotted,
blue line), $|C_{gg-}C_{\gamma\gamma -}|^2$ (narrow, dashed, orange
line), $|C_{gg\tilde t\tilde t}C_{\gamma\gamma\tilde t\tilde t}|^2$
(solid, black line), and $|C_{ggH}C_{\gamma\gamma H}|^2$ (dash-dotted, black
line). The right panel shows cross sections times branching ratios 
into $\gamma\gamma$: contribution in the narrow-resonance approximation
of the larger-mass eigenstate (dotted, blue line), contribution in the
narrow-resonance approximation of the smaller-mass eigenstate (narrow,
dashed, orange line), sum of the contributions in the narrow-resonance
approximation (thick, dashed, red line), exact cross section times 
branching ratio into $\gamma\gamma$ (thin, black line), and exact cross
section times branching ratio into $\gamma\gamma$ in the absence of
mixing (dash-dotted, black line).}
\end{figure}

As can be seen from the left panel of Fig.~\ref{fig:BR-8n}, the effect
of mixing on the short-distance coefficients now produces only monotonic 
functions with no peaks or dips. 
This simple structure is attributable to the fact that $C_{H gg}$ and
$C_{H \gamma\gamma}$  are dominated by their imaginary parts.
Furthermore, as can be seen from the figure, the product of Higgs
short-distance coefficients is much larger in magnitude than the product
of stop-antistop coefficients, and so the Higgs coefficients dominate at
minimal mixing. At $m_H=720$~GeV and $m_H=780$~GeV, the effects on the
short-distance coefficients are still considerable, especially at
$m_H=780$~GeV.The stop-antistop width $\Gamma_{\tilde t\tilde t}$ at
$m_H = 720$~GeV is about 0.24~GeV and increases as $m_H$ approaches the
stop-antistop threshold. Therefore, owing to the dominance of the Higgs
short-distance coefficients, the stop-antistop resonance does not
contribute greatly to the cross section times the branching ratio.  At
maximal mixing, there are two resonances, whose widths are equal to
about $\tfrac{1}{2}\Gamma_H$ and whose values of $|C_{gg
\pm}C_{\gamma\gamma \pm}|^2$ are about
$\tfrac{1}{2}|C_{ggH}C_{\gamma\gamma H}|^2$. Consequently, the cross
section times the branching ratio changes very little from minimal mixing
to maximal mixing, as can be seen in the right panel of
Fig.~\ref{fig:BR-8n}. Even at $m_H=720$~GeV and $m_H=780$~GeV, the
values of the cross section times the branching ratio lie somewhat above the
values in the absence of mixing. The result from the narrow-resonance
approximation for the cross section is essentially featureless. We note
that it agrees well with the exact result.

\subsubsection{$\kappa=8$, $\Gamma_{\tilde t}=0.1$~GeV}

Finally, we consider the case of strong Higgs-stop-antistop coupling, 
$\kappa=8$, and large stop width, $\Gamma_{\tilde t}=0.1$~GeV.

\begin{figure}
\epsfig{file=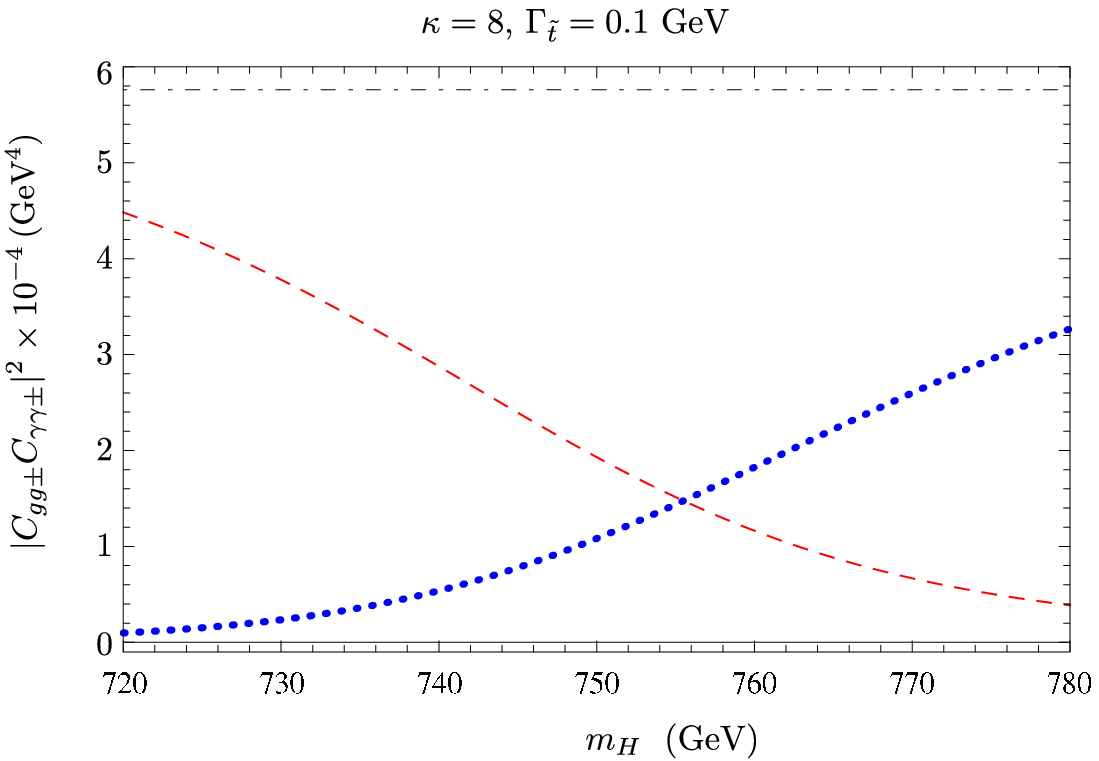,width=0.45\columnwidth} 
\epsfig{file=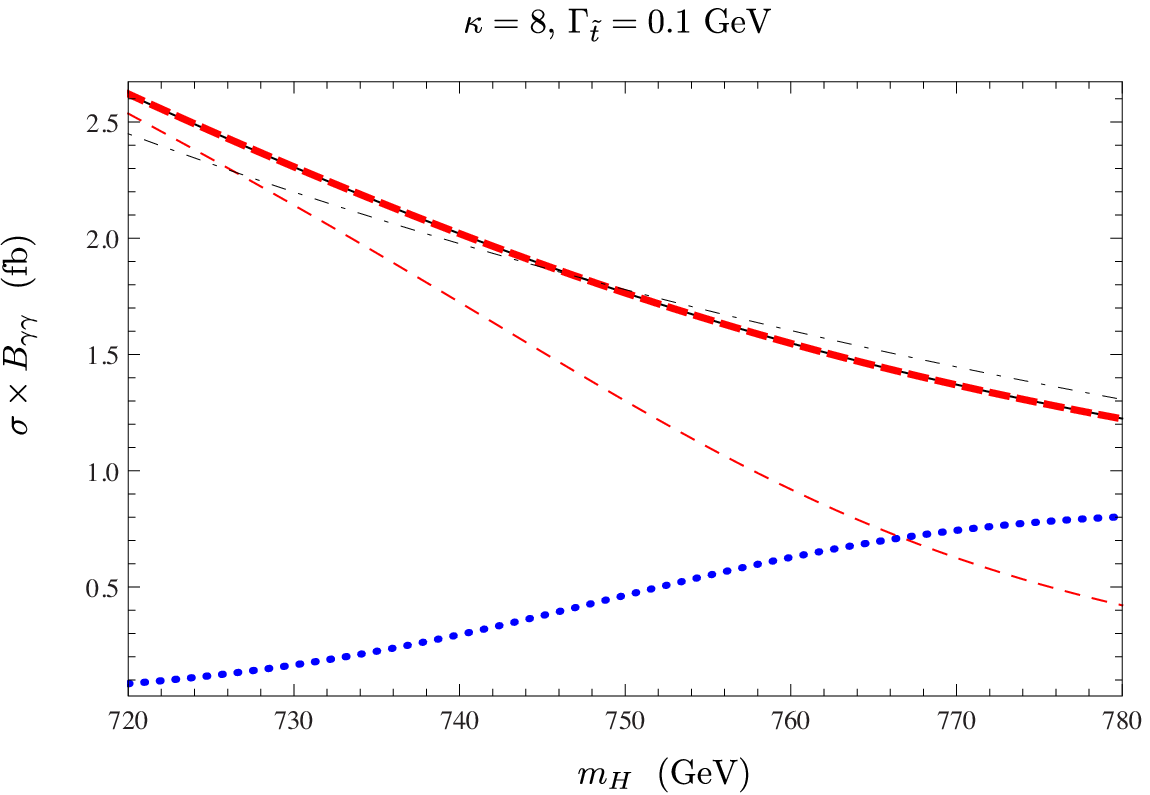,width=0.45\columnwidth}
\caption{\label{fig:BR-8w}%
Numerical results for the case in which the stop-antistop Green's
function is given by a Breit-Wigner resonance for $\kappa=8$ and
$\Gamma_{\tilde t}=0.1$~GeV. Quantities are presented as functions of
$m_H$. The left panel shows the products of absolute squares of
short-distance coefficients: $|C_{gg+}C_{\gamma\gamma +}|^2$ (dotted,
blue line), $|C_{gg-}C_{\gamma\gamma -}|^2$ (narrow, dashed, orange
line), $|C_{gg\tilde t\tilde t}C_{\gamma\gamma\tilde t\tilde t}|^2$
(solid, black line), and $|C_{ggH}C_{\gamma\gamma H}|^2$ (dash-dotted, black
line). The right panel shows cross sections times branching ratios 
into $\gamma\gamma$: contribution in the narrow-resonance approximation
of the larger-mass eigenstate (dotted, blue line), contribution in the
narrow-resonance approximation of the smaller-mass eigenstate (narrow,
dashed, orange line), sum of the contributions in the narrow-resonance
approximation (thick, dashed, red line), exact cross section times 
branching ratio into $\gamma\gamma$ (thin, black line), and exact cross
section times branching ratio into $\gamma\gamma$ in the absence of
mixing (dash-dotted, black line).}
\end{figure}

As can be seen from the left panel of Fig.~\ref{fig:BR-8w}, the effect
of mixing on the short-distance coefficients is essentially the same as 
in the case of $\kappa=8$ and $\Gamma_{\tilde t}=0.1$~MeV 
(Fig.~\ref{fig:BR-8n}).
Again, owing to the dominance of the imaginary parts of  $C_{H gg}$ and
$C_{H \gamma\gamma}$, the effects of mixing on the short-distance
coefficients result in a simple structure. Since the product of Higgs
short-distance coefficients is much larger in magnitude than the product
of stop-antistop coefficients,  the Higgs coefficients dominate at
minimal mixing. The stop-antistop width at $m_H = 720$~GeV is now
$\Gamma_{\tilde t\tilde t} \approx 0.44$~GeV. This larger stop-antistop
width has only a small effect on the cross section times branching
ratio at minimal mixing, where the Higgs contribution is dominant, and
changes the resonance widths only slightly at maximal mixing. 
Consequently, as in the case of $\kappa=8$ and $\Gamma_{\tilde
t}=0.1$~MeV, the cross section times branching ratio changes very
little from minimal mixing to maximal mixing. This can be seen in the
right panel of Fig.~\ref{fig:BR-8n}. In this case, the contributions
from the two eigenstates sum to produce a total of the cross section times
the branching ratio that deviates very little from the total of the 
cross section times
the branching ratio in the the absence of mixing. Again, the result from
the narrow-resonance approximation for the cross section is essentially
featureless, and it agrees well with the exact result.

\subsection{Differences between the Breit-Wigner resonance and the 
Coulomb-Schr\"odinger Green's function\label{sec:differences}}

As we will see, many of the qualitative features of the model in which
we replace the stop-antistop Green's function with a Breit-Wigner
resonance persist when we model the stop-antistop Green's function with
the Coulomb-Schr\"odinger Green's function. There are, however, several
important differences.

First, the  Coulomb-Schr\"odinger Green's function develops an imaginary
part above the stop-antistop threshold, owing to the fact that the
physical states can decay into a stop-antistop pair. In the case of
large Higgs-stop-antistop coupling, this imaginary part can broaden
the higher-mass physical state significantly and lead to a substantial
reduction in its contribution to the cross section.

Second, the logarithm of $\tilde{E}$ in Eq.~(\ref{G-C-S}) can produce 
additional structure near the stop-antistop threshold. For the 
examples that we have considered, this additional structure appears to 
have a small effect on the cross section times the branching ratio to 
$\gamma\gamma$.

Third, the Coulomb-Schr\"odinger Green's function contains multiple
bound-state poles. As we will see, the additional poles beyond the
ground-state pole do not have a dramatic effect on the cross section
times the branching ratio to
$\gamma\gamma$.

Fourth, the Coulomb-Schr\"odinger Green's function has a much more
complicated dependence on $\hat{s}$ than does the Breit-Wigner
resonance. This situation is analyzed in detail in 
Appendix~\ref{sec:C-S-diagonal}.
 The generalization of the matrix in
Eq.~(\ref{matrix-amps-one-pole}) is diagonalized by a similarity
transformation that depends on $\hat{s}$. Each of the matrix eigenvalues
$\alpha_\pm$ can have multiple poles in its inverse. The real parts of
the poles locations, $m_\pm$, and the imaginary parts of the pole
locations, $m_\pm \Gamma_\pm$, determine the resonance widths, which are
different than in the Breit-Wigner case.  An important additional
difference is that the residues of those poles, $Z_\pm$, which are equal
to unity in the Breit-Wigner case, can now have magnitudes that are
larger or smaller than unity. This change in the pole residues is driven
largely by the term that is proportional to $1/\lambda$ in
Eq.~(\ref{G-C-S}). However, the specific value of the residue depends on
the other terms in Eq.~(\ref{G-C-S}), as well.

Fifth, in the Coulomb-Schr\"odinger analysis, we include the effects
of the $t$-channel Higgs exchange on the stop-antistop propagator. Had
we included these effects in the Breit-Wigner analysis, they would have
produced only small shifts in the mass of the Breit-Wigner resonance. In
the Coulomb-Schr\"odinger analysis, it turns out that the $t$-channel
Higgs exchange has a negligible effect for the case $\kappa=1$. However,
it produces small, but noticeable, effects for the case $\kappa=8$,
particularly for the smaller stop width. These
$t$-channel-Higgs-exchange effects do not change the qualitative picture
for the total cross section.

As we have mentioned, in the approximation in which a resonance
amplitude is given by a Breit-Wigner form and the separation of the
resonance from other resonances is small in comparison to the resonance
widths, the contribution of a resonance to the cross section is
proportional to the square of the maximum height of the absolute value
of the amplitude times the full width at half maximum of the peak in the
absolute value of the amplitude. Specifically, the
cross-section contribution is proportional to $|Z_\pm|^2/\Gamma_\pm$.

\section{Case of the Coulomb-Schr\"odinger Green's
function\label{sec:numerical}}

In this section, we present numerical results for the $gg\to
\gamma\gamma$ amplitudes and the associated LHC cross sections at 
$\sqrt{s}=13$~TeV for the case in which the stop-antistop Green's 
function is calculated from the Coulomb-Schr\"odinger 
Green's function [Eq.~(\ref{G-C-S})].

Heavy-Higgs production and decay rates depend not only on the Higgs
coupling to the stop squark, but also on the Higgs couplings to the top
and bottom quarks. As we have mentioned, we make use of an
intermediate value of $\tan\beta$, setting $\tan\beta=\sqrt{m_t/m_b}$,
where $m_t$ and $m_b$ are the $\overline{\rm MS}$ running masses at the
scale $2 m_{\tilde t}$. At this value of $\tan\beta$, the Higgs decay
width into third generation fermions is minimized and is about 1.2~GeV.
The additional contributions to the Higgs production rate and decay
width that are associated with Higgs couplings to stop-antistop pairs
are automatically taken into account within our theoretical framework.
The contributions to the total width from the decay of the heavy Higgs boson 
into pairs of Higgs bosons or heavy gauge bosons tend to be small, and we omit 
them in our analysis.

Our results depend strongly on the Higgs coupling to the stop-antistop
pair and on the stop width. Therefore, we present numerical results for
two representative cases of weak and strong Higgs coupling to the
stop-antistop pair and for small and moderate values of the stop width.
In particular, as in our analysis of the Breit-Wigner-resonance
model, we give results for the cases $\kappa = 1$ and $\Gamma_{\tilde
t} = 0.1$ MeV, $\kappa = 1$ and $\Gamma_{\tilde t} = 0.1$ GeV, $\kappa =
8$ and $\Gamma_{\tilde t} = 0.1$ MeV, and $\kappa = 8$ and
$\Gamma_{\tilde t} = 0.1$ GeV. We use the input values of the various
parameters that were discussed in
Sec.~\ref{sec:breit-wigner-qualitative}. We present results for various
values of the Higgs mass and the partonic center-of-mass energy
$\sqrt{\hat{s}}$.

In computing cross sections, we take into account only the gluon-gluon
initiated process, which has been the focus of our discussion. As we
have mentioned, the true stop-antistop Green's function likely contains
only a few bound states below threshold. Therefore, we give results that
are obtained by taking into account only the first term or the first
three terms in the sum in Eq.~(\ref{G-C-Sa}). For comparison, we also give
results that are based on the full Coulomb-Schr\"odinger Green's
function in Eq.~(\ref{G-C-Sa}). In the figures for the amplitudes
below, we show, for clarity, the results that are obtained by retaining
only one pole in the stop-antistop Green's function. That is, take only
the $n=1$ term in the sum in Eq.~(\ref{G-C-Sa}). In these figures, we
show the following : (1) $|A_{\rm tot}|$; (2) $|A^{\rm
bare}_H|=|A_1^{\rm bare}+A_2^{\rm bare}+A_3^{\rm bare}+ A_4^{\rm
bare}|$, where superscript ``bare'' means that the stop-antistop
corrections to the Higgs propagator in Eq.~(\ref{A1}), which are
proportional to $C_{H\tilde t\tilde t}^2$, have been neglected; and (3)
$|A_{\tilde t\tilde t}^{\rm bare}|$, where $A_{\tilde t\tilde t}^{\rm
bare} = A_5$ is the stop-antistop amplitude in the absence of Higgs
coupling to the stop.
Note that $A^{\rm bare}_H$ contains all of the  Higgs-form-factor
contributions that were discussed in Sec.~\ref{sec:form-factors}.
However, the absence of stop-antistop corrections to the Higgs
propagator in $A^{\rm bare}_H$ affects that amplitude in two important
ways: (1) the Higgs-stop-antistop-mixing effects that lead to the
displacement of the physical mass eigenvalues from threshold are not
present and (2) some of the corrections to the Higgs width 
that are associated with Higgs decays into stop-antistop pairs for Higgs
masses above the stop-antistop threshold are not present.

We remind the reader that, as we have explained in
Sec.~\ref{sec:breit-wigner}, the cross-section contribution of a
resonance whose amplitude can be approximated by a Breit-Wigner form is
proportional, in the narrow-width approximation, to the square of the
maximum height of the absolute value of the amplitude times the full width
at half maximum of the absolute value of the amplitude.

\subsection{$\kappa = 1$, $\Gamma_{\tilde t} = 0.1$ MeV}

In this case, the Higgs boson couples only weakly to the  
stop, and the stop width is much less than the Higgs width.

\begin{figure}
\epsfig{file=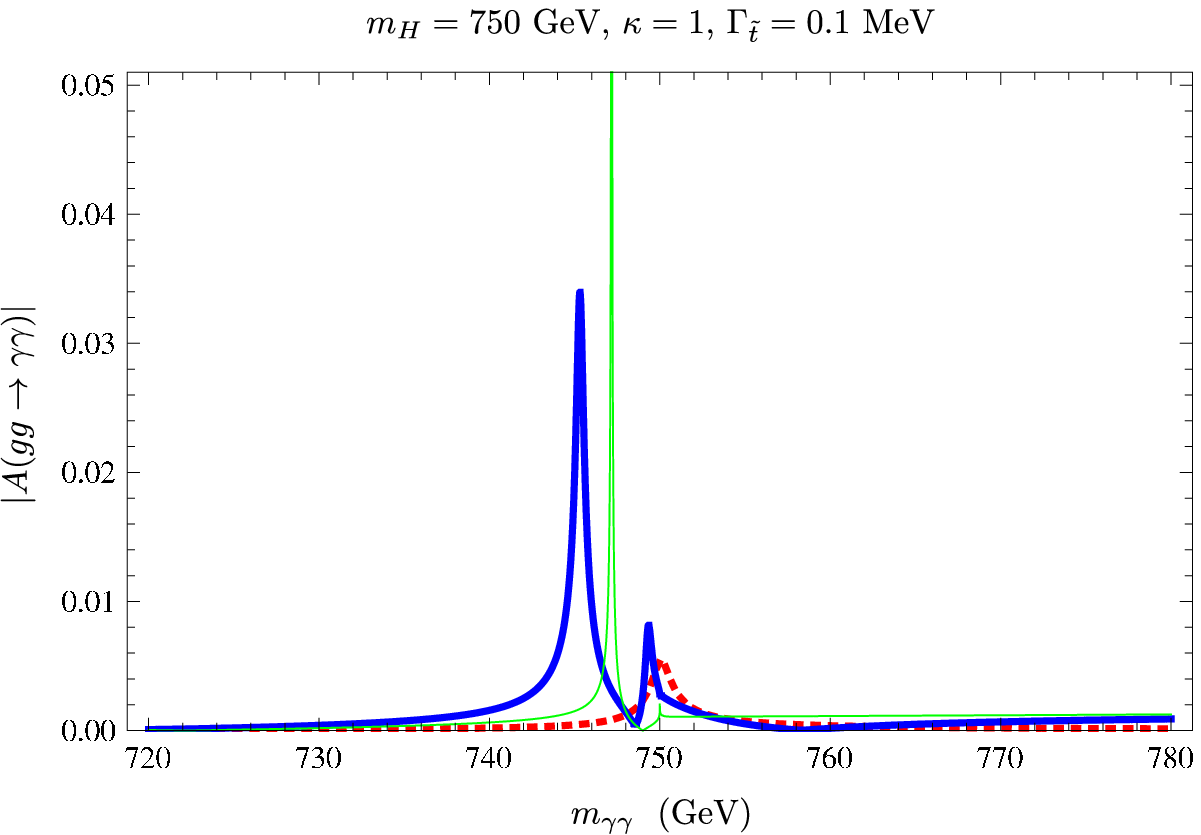,width=0.45\columnwidth} 
\epsfig{file=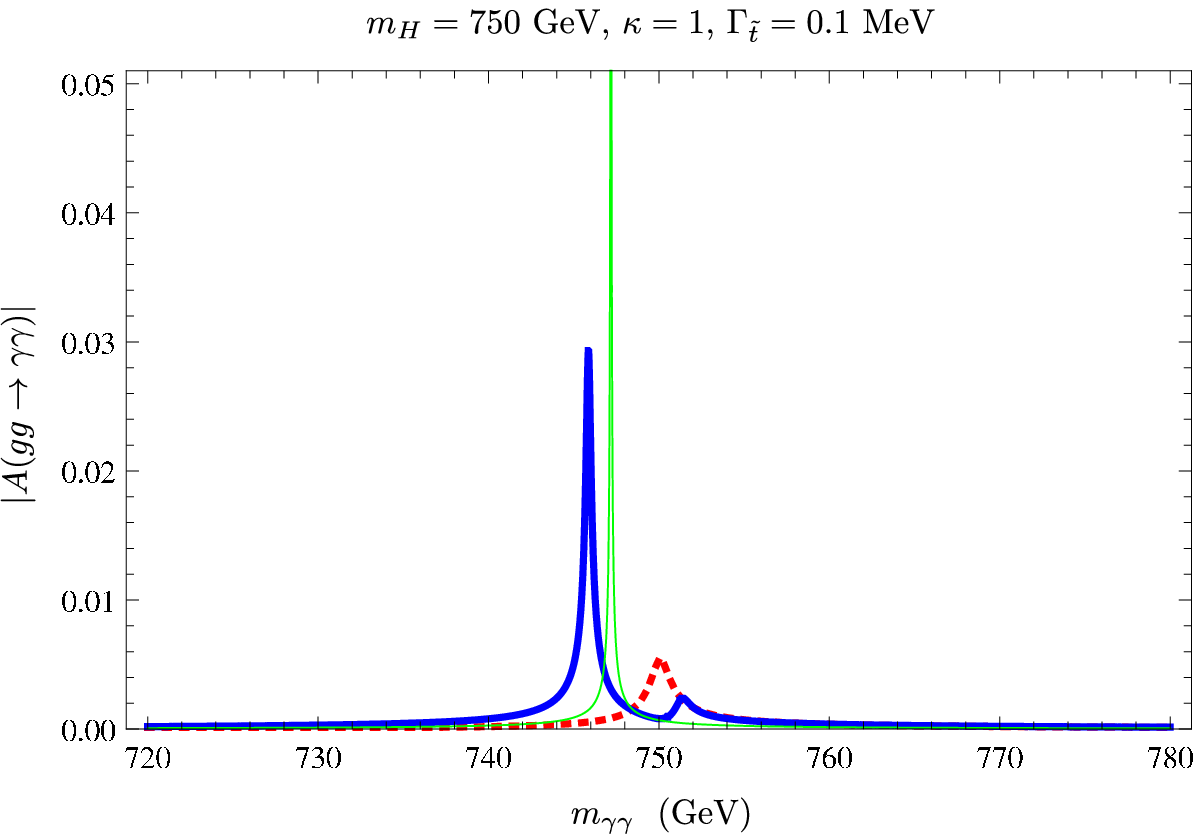,width=0.45\columnwidth}
\epsfig{file=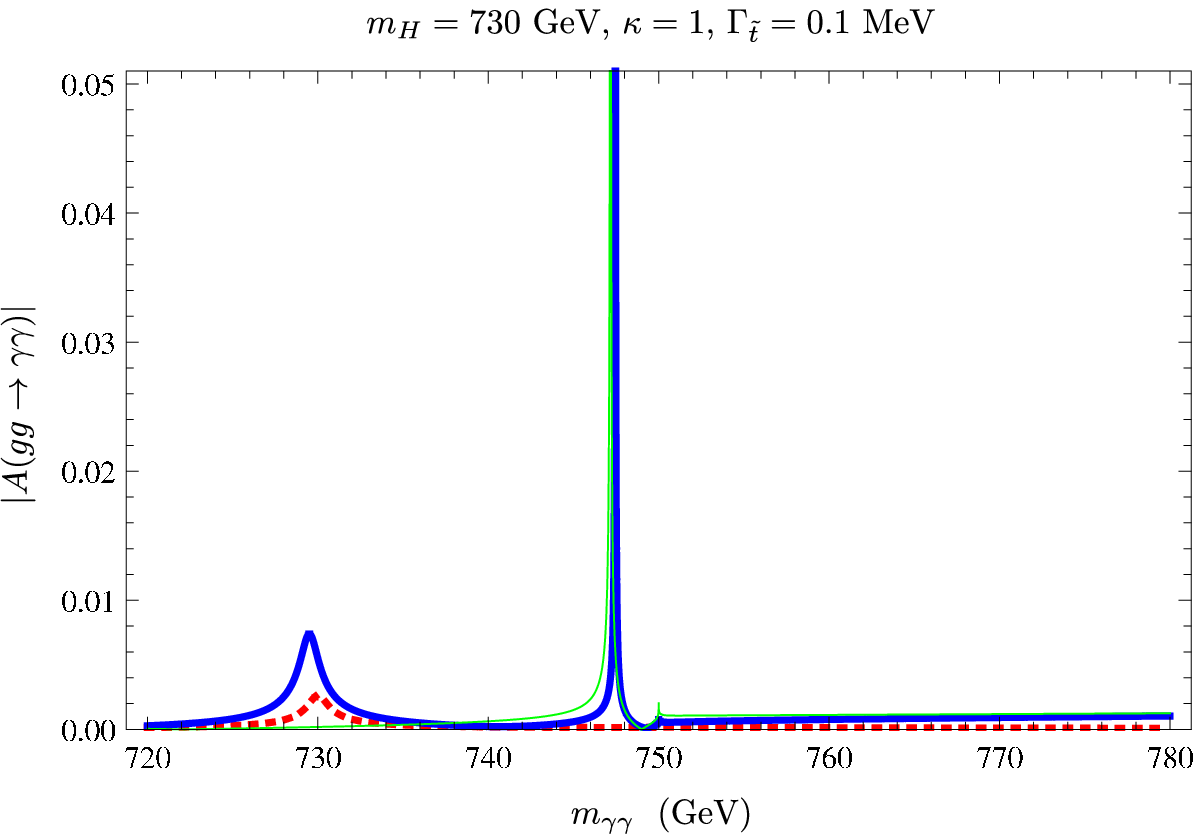,width=0.45\columnwidth} 
\epsfig{file=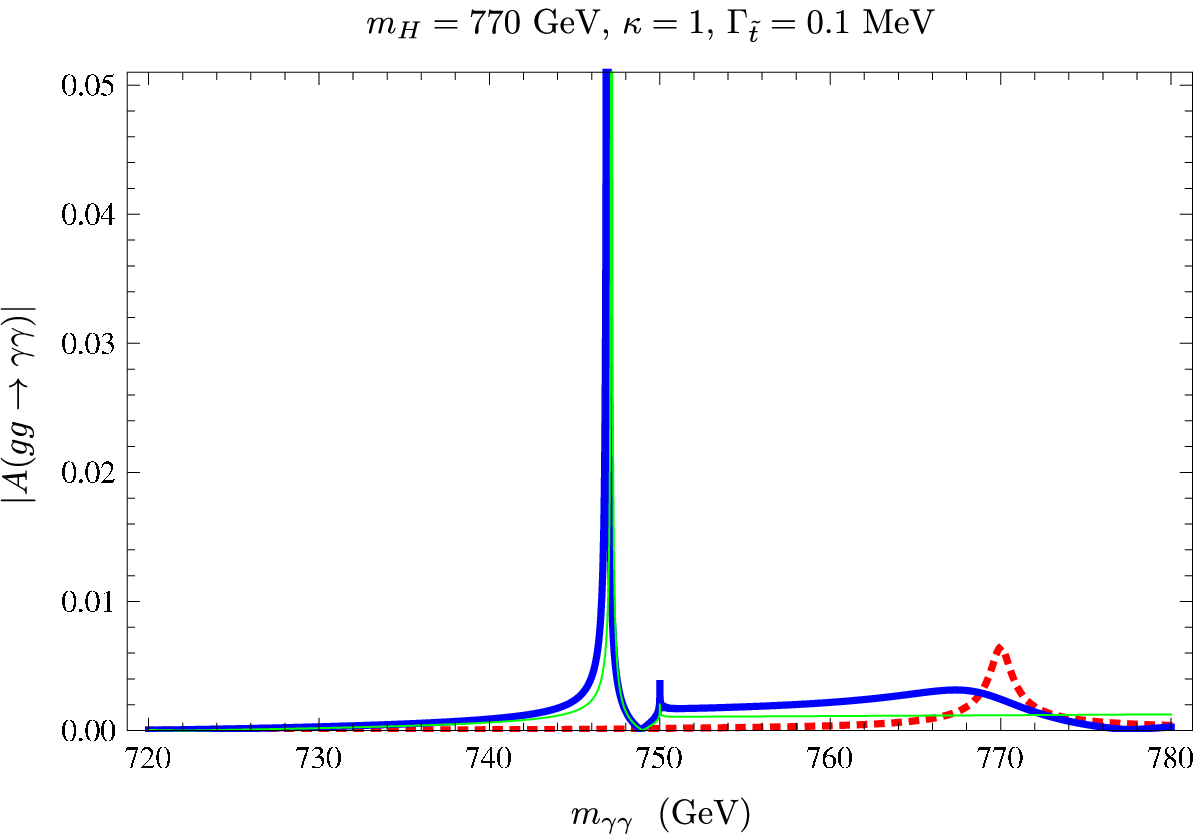,width=0.45\columnwidth}
\caption{\label{fig:1Mamp}%
Amplitudes for the case $\kappa=1$, $\Gamma_{\tilde t}=0.1$~MeV.
Upper left figure: the amplitudes $|A_{\rm tot}|$ (thick, blue line),
$|A_{\tilde t\tilde t}^{\rm bare}|$ (thin, green line), and $|A_H^{\rm
bare}|$ (dashed, red line) vs $m_{\gamma \gamma}$ for $m_H =
750$~GeV. Upper right figure: the same amplitudes, but with the
stop-antistop propagator replaced with a Breit-Wigner resonance, as in
Eq.~(\ref{Breit-Wigner}).  Lower left figure:
the same as the upper left figure, but with $m_H=730$~GeV. Lower right
figure: the same as the upper left figure, but with $m_H=770$~GeV.
}
\end{figure}

In Fig.~\ref{fig:1Mamp}, we show $|A_{\rm tot}|$, $|A^{\rm bare}_H|$,
and $|A_{\tilde t\tilde t}^{\rm bare}|$  for $m_H = 720$, 750, and 780
GeV. The results for the three different Higgs masses show that, for
$\kappa = 1$, the mixing has a small effect on the physical masses,
which remain close to their values in the absence of mixing. At $m_H =
720$~GeV we  can clearly identify the Higgs and stoponium contributions
to the total amplitude. At $m_H = 780$~GeV, we can also see the separate
contributions from the Higgs and the stoponium peaks, but we see a large
increase in the width of the Higgs peak that is associated with the
Higgs decay to a stop-antistop pair. At $m_H = 750$~GeV, where the
mixing is maximal, the physical masses are slightly displaced relative
to the Higgs and stoponium masses. Comparison of the upper left panel
with the upper right panel shows that $|A_{\rm tot}|$ for the
Coulomb-Schr\"odinger Green's function and the Breit-Wigner Green's
function are quite similar, although the lower-mass peak is somewhat
broader and higher in the Coulomb-Schr\"odinger case. Note that, at
maximal mixing, both physical peaks are much broader than the
unmixed stoponium peak.

\begin{figure}
\epsfig{file=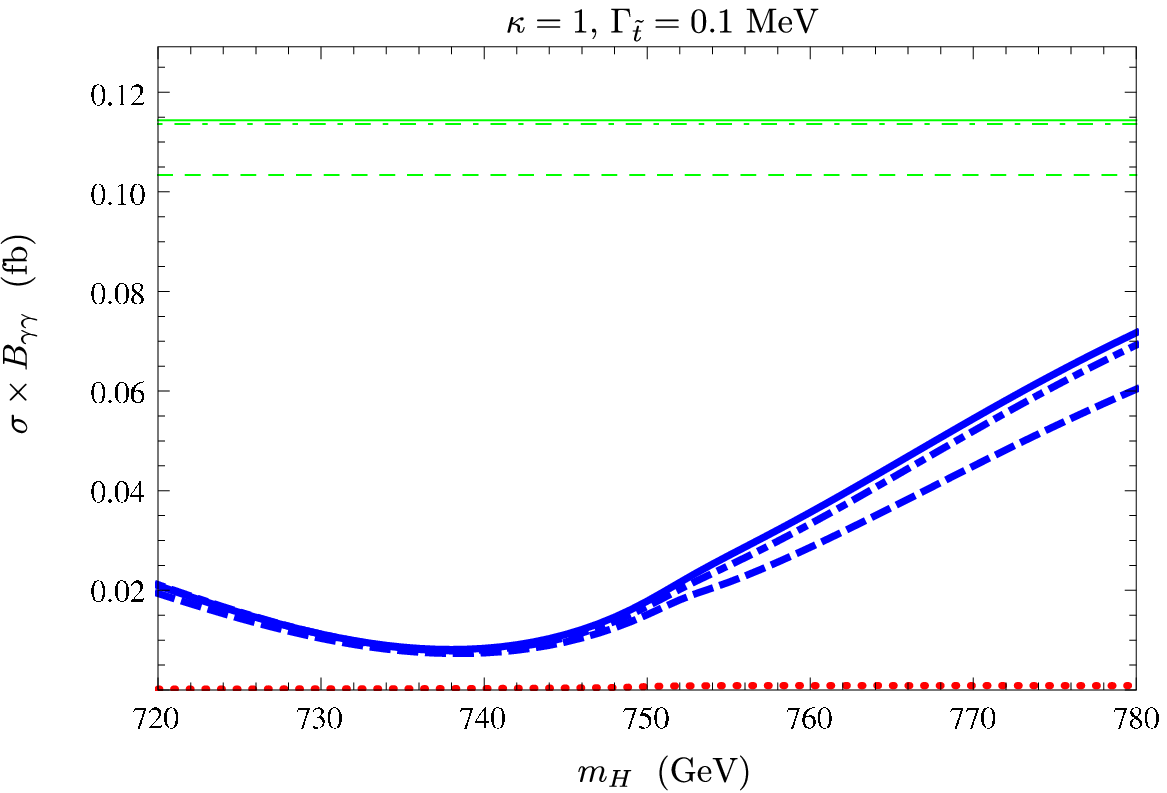,width=10cm}
\caption{\label{fig:1MCS}%
Cross sections for the case $\kappa=1$, $\Gamma_{\tilde 
t}=0.1$~MeV: $\sigma_{\rm tot}$ (thick, blue lines),
$\sigma_H^{\rm bare} $ (dotted, red line), and $\sigma_{\tilde t
\tilde t}^{\rm bare}$ (thin, green lines) vs $m_H$. In the
cases of $\sigma_{\rm tot}$ and $\sigma_{\tilde t \tilde t}^{\rm
bare}$, the dashed, dashed-dotted, and solid lines correspond,
respectively, to taking $1$, $3$, or all terms in the sum in
Eq.~(\ref{G-C-Sa}).
}
\end{figure}

In Fig.~\ref{fig:1MCS}, we show the total diphoton production cross
section $\sigma_{\rm tot}$ as a function of $m_H$. For comparison, we
also show $\sigma_H^{\rm bare}$ (which corresponds to $A_H^{\rm
bare}$) and $\sigma_{\tilde t \tilde t}^{\rm bare}$ (which corresponds
to $A_{\tilde t\tilde t}^{\rm bare}$) as functions of $m_H$. 
Figure~\ref{fig:1MCS} shows that there are only small quantitative changes
in $\sigma_{\rm tot}$ as one includes additional stoponium poles in the
stop-antistop Green's function.

A comparison of Fig.~\ref{fig:1MCS} with the right panel of
Fig.~\ref{fig:BR-1n} shows that the total cross section $\sigma_{\rm
tot}$ in the Coulomb-Schr\"odinger case has the same qualitative
features as in the Breit-Wigner case.  The  Higgs cross section $\sigma_H^{\rm bare}$ is much less than the
stop-antistop cross section $\sigma_{\tilde t \tilde t}^{\rm bare}$. The
cross section is dominated by the width of the narrowest peak. At
minimal mixing, this narrowest peak corresponds to the stoponium peak.
At maximal mixing, the width of the narrowest physical peak is much
greater than at minimal mixing, and the height shrinks roughly as the 
inverse of the width, resulting in a suppression of the cross
section. As in the Breit-Wigner case (Sec.~\ref{sec:breit-wigner-1n}),
this suppression is so great that it overwhelms the peaking effect that
results from mixing of the short-distance coefficients. We see that,
even at $m_H=720$~GeV and $m_H=780$~GeV, mixing broadens the narrowest
peak and reduces its height sufficiently that the total cross
section is well below $\sigma_{\tilde t \tilde t}^{\rm bare}$.

\subsection{$\kappa = 1$, $\Gamma_{\tilde t} = 0.1$ GeV}

In this case, the Higgs boson still couples weakly to the stop, but the 
stop width is much closer to the Higgs width than in the previous 
example.

\begin{figure}
\epsfig{file=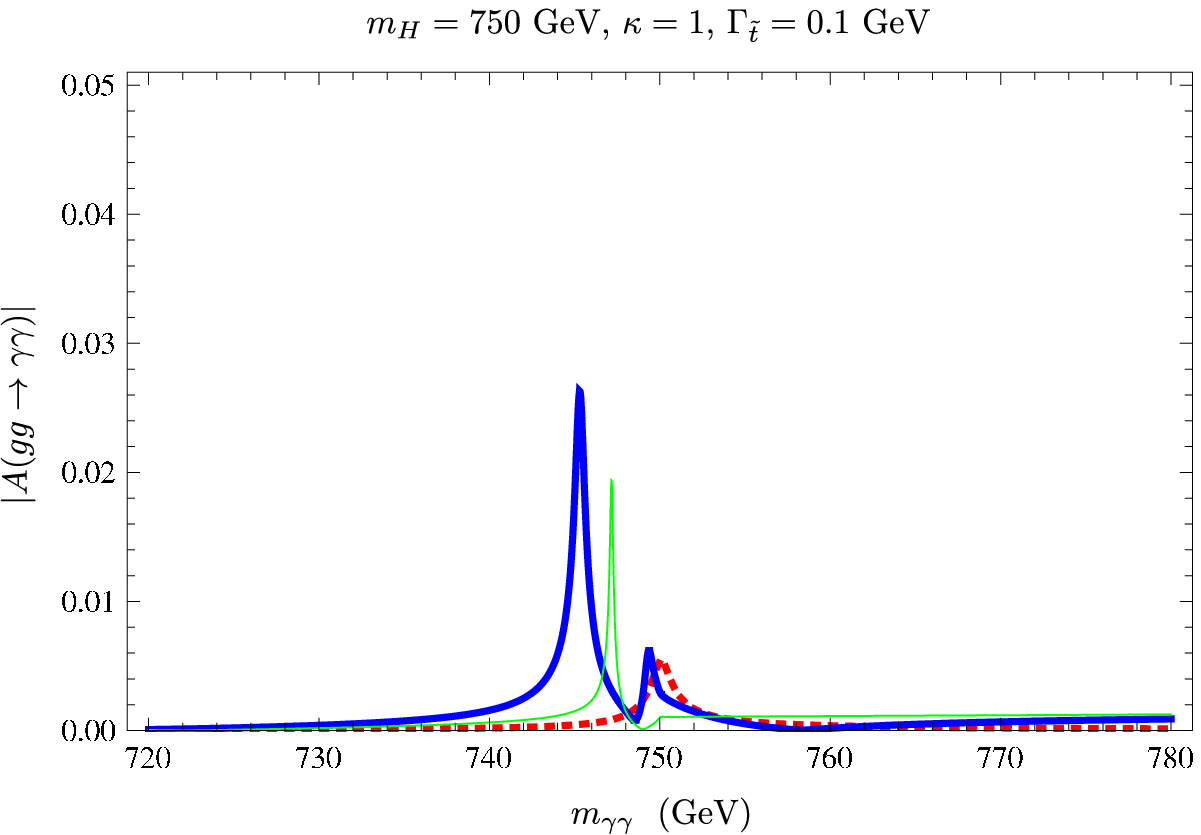,width=0.45\columnwidth}
\epsfig{file=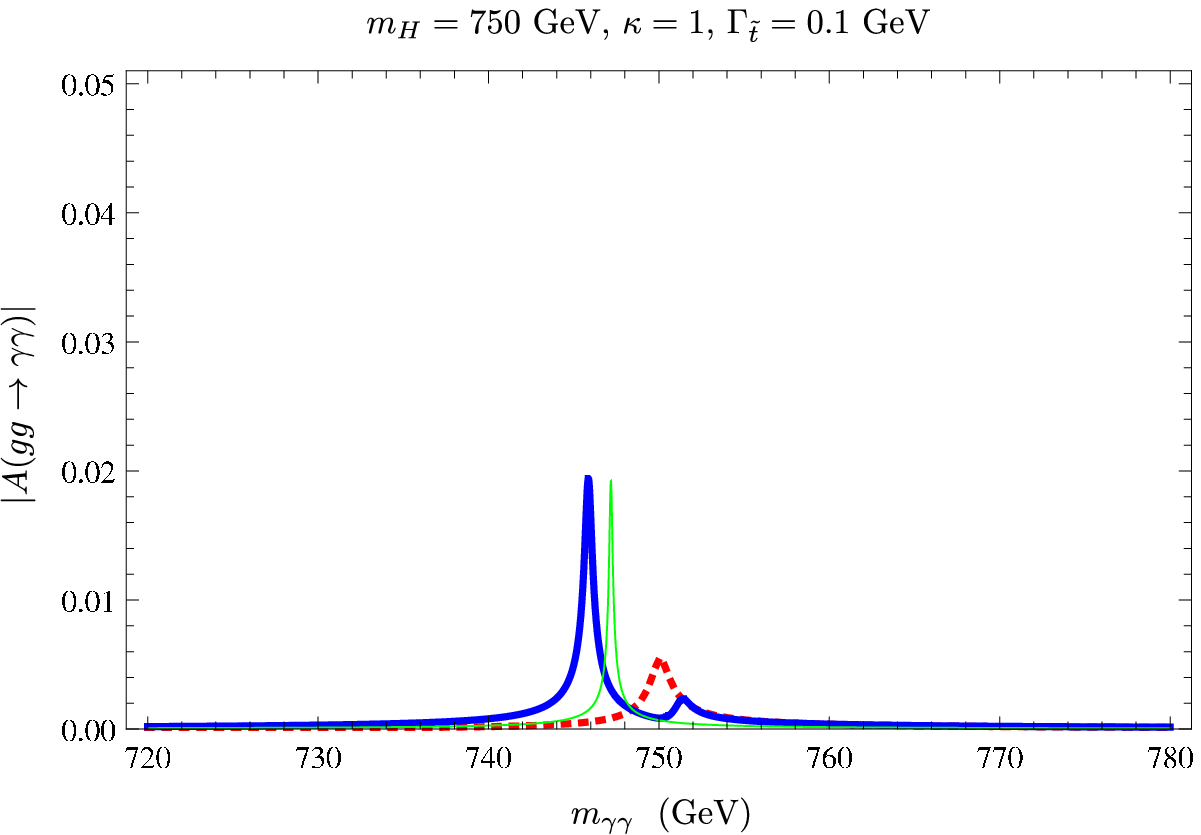,width=0.45\columnwidth} 
\caption{\label{fig:1Gamp}%
Amplitudes for the case $\kappa=1$, $\Gamma_{\tilde t}=0.1$~GeV.
Left figure: the amplitudes $|A_{\rm tot}|$ (thick, blue line),
$|A_{\tilde t\tilde t}^{\rm bare}|$ (thin, green line), 
and $|A_H^{\rm bare}|$ (dashed, red line) vs $m_{\gamma \gamma}$ for
$m_H = 750$~GeV. Right figure: the same amplitudes, but with the
stop-antistop propagator replaced with a Breit-Wigner resonance, as in
Eq.~(\ref{Breit-Wigner}).
}
\end{figure}

In Fig.~\ref{fig:1Gamp}, we show  $|A_{\rm tot}|$, 
$|A^{\rm bare}_H|$, and $|A_{\tilde t\tilde t}^{\rm bare}|$. 
The left and right panels show the results that are obtained in the 
Coulomb-Schr\"odinger case and the Breit-Wigner case, respectively.
Again, these two choices for the stop-antistop Green's function result
in qualitatively similar amplitudes. However, the height and width
of the lower-mass resonance are both enhanced in the
Coulomb-Schr\"odinger case relative to the Breit-Wigner case. At
$m_H=750$~GeV, where mixing is maximal, we see that the physical masses
are displaced slightly relative to the Higgs and stoponium masses.
Comparison  with Fig.~\ref{fig:1Mamp} shows that there is a marked
increase in the width of the stoponium peak in $A_{\tilde t\tilde
t}^{\rm bare}$, owing to the increase in the stop width. Because of
mixing effects, the width of the narrowest peak in $A_{\rm tot}$ is much
larger than the width of the stoponium peak in $A_{\tilde t\tilde
t}^{\rm bare}$. The width of the narrowest peak in $A_{\rm tot}$  is
larger in Fig.~\ref{fig:1Gamp} than in Fig.~\ref{fig:1Mamp}, owing to
the increase in the width of the stop.

\begin{figure}
\epsfig{file=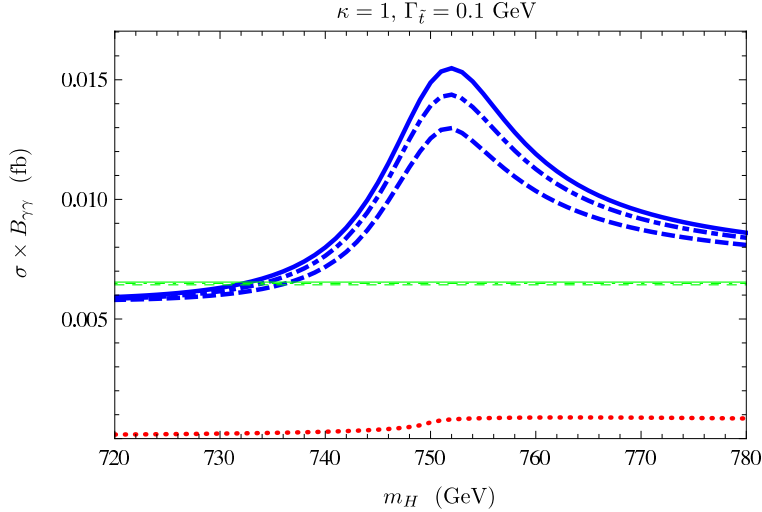,width=10cm}
\caption{\label{fig:1GCS}%
Cross sections for the case $\kappa=1$, $\Gamma_{\tilde t}=0.1$~GeV:
$\sigma_{\rm tot}$ (thick, blue lines),
$\sigma_H^{\rm bare} $ (dotted, red line), and $\sigma_{\tilde t
\tilde t}^{\rm bare}$ (thin, green lines) vs $m_H$. In the
cases of $\sigma_{\rm tot}$ and $\sigma_{\tilde t \tilde t}^{\rm
bare}$, the dashed, dashed-dotted, and solid lines correspond,
respectively, to taking $1$, $3$, or all terms in the sum in
Eq.~(\ref{G-C-Sa}).
}
\end{figure}

In Fig.~\ref{fig:1GCS}, we display  $\sigma_{\rm tot}$,
$\sigma_H^{\rm bare} $, and $\sigma_{\tilde t \tilde t}^{\rm
bare}$, as functions of $m_H$. As can be seen, there are only small
quantitative differences that are associated with the inclusion of
additional stoponium poles in the stop-antistop Green's function.

A comparison of Fig.~\ref{fig:1GCS} with the right panel of
Fig.~\ref{fig:BR-1w} shows that the total cross section $\sigma_{\rm
tot}$ in the Coulomb-Schr\"odinger case again has the same qualitative
features as in the Breit-Wigner case. However, there are minor
differences in the shapes, and the Coulomb-Schr\"odinger cross section
is considerably enhanced relative to the Breit-Wigner cross section,
owing to the increase in the height and width of the lower-mass
resonance, which can be seen in Fig.~\ref{fig:1Gamp}. Once again, the
Higgs cross section $\sigma_H^{\rm bare}$ is
much less than the stop-antistop cross section $\sigma_{\tilde t \tilde
t}^{\rm bare}$, although $\sigma_{\tilde t \tilde t}^{\rm bare}$ is
reduced in this broad-stop-width case in comparison to $\sigma_{\tilde t
\tilde t}^{\rm bare}$ in the small-stop-width case. Nevertheless, the
cross section is dominated by the width of the narrowest peak, which
corresponds, at minimal mixing, to the stoponium peak. We see that, in
contrast with the small-stop-width case, the cross section now displays
a peak at maximal mixing. As we explained in
Sec.~\ref{sec:numerical-1n}, this peak arises from the effects of mixing
on the short-distance coefficients and is unrelated to
threshold-enhancement effects. A peak persists in $\sigma_{\rm tot}$
because, owing to the large stop width, the width of the narrowest peak
in $A_{\rm tot}$ does not change sufficiently between minimal and
maximal mixing to reverse the peaking effect from the
short-distance coefficients.

\subsection{$\kappa = 8$, $\Gamma_{\tilde t} = 0.1$ MeV}

Next, we discuss the case of a large Higgs-stop-antistop coupling, 
$\kappa=8$, and a small stop width, $\Gamma_{\tilde t}=0.1$~MeV.

Large values of the Higgs-stop-antistop coupling have a very large impact
on the diphoton production rate, as has been emphasized in the context 
of perturbative calculations in Ref.~\cite{Djouadi:2016oey}. As
we will see, for $\kappa = 8$, the Higgs-stop-antistop mixing effects
become dramatic, and it is essential to include those effects, which
go beyond the effects that are contained in fixed-order perturbation
theory, in order to compute the diphoton rate reliably.

\begin{figure}
\epsfig{file=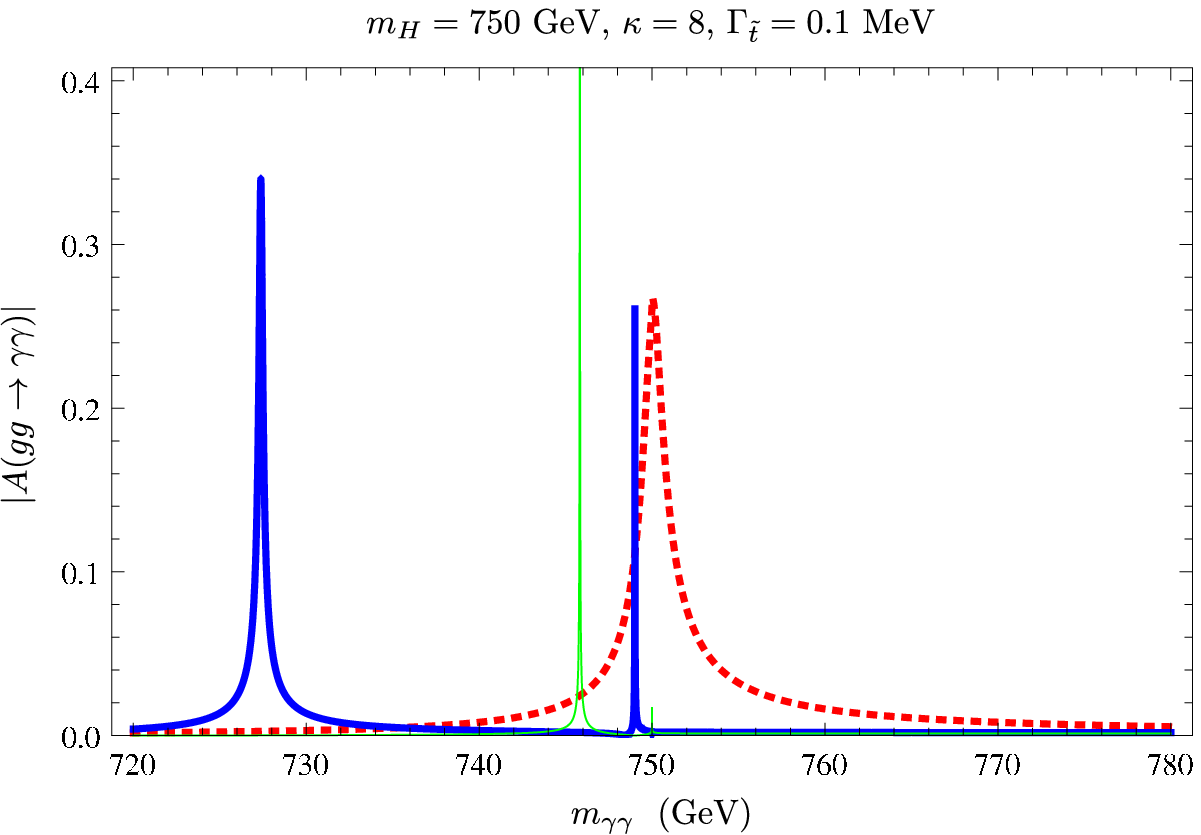,width=0.45\columnwidth} 
\epsfig{file=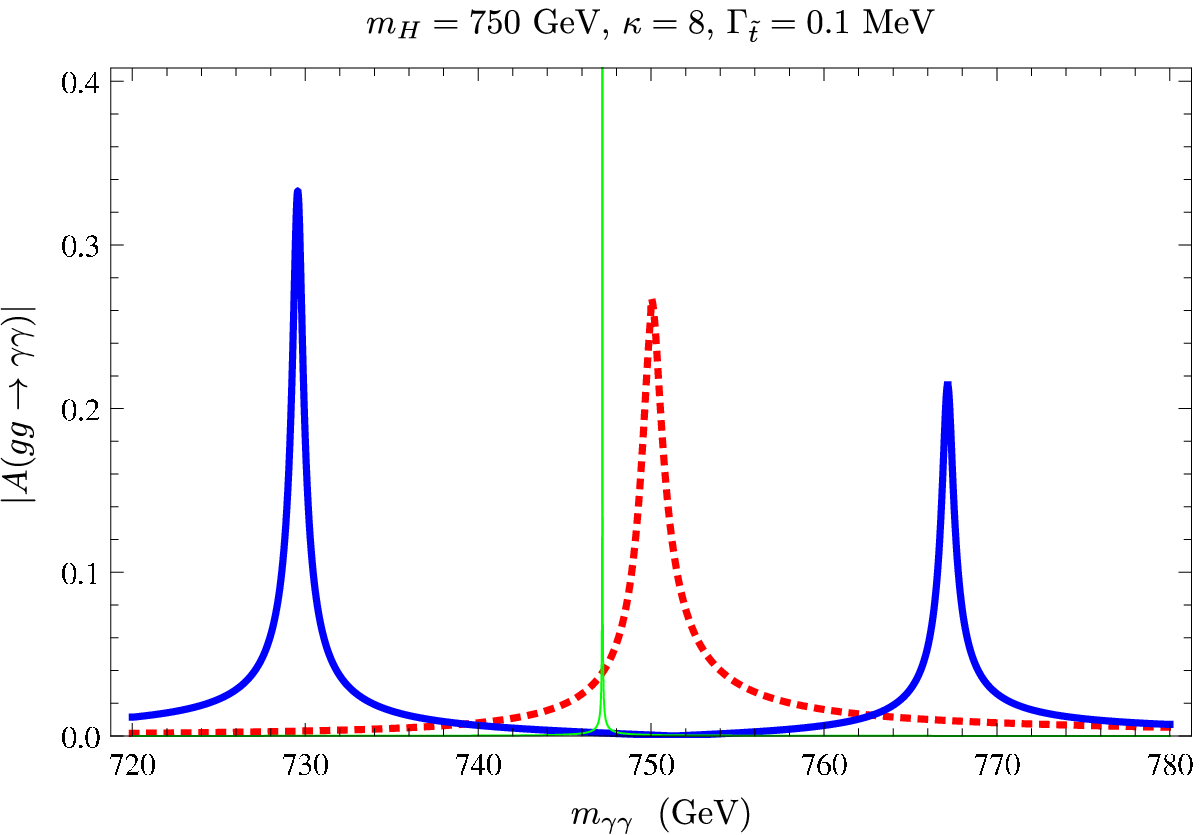,width=0.45\columnwidth}
\caption{\label{fig:8Mamp}%
Amplitudes for the case $\kappa=8$, $\Gamma_{\tilde t}=0.1$~MeV.
Left figure: the amplitudes $|A_{\rm tot}|$ (thick, blue line),
$|A_{\tilde t\tilde t}^{\rm bare}|$ (thin, green line), 
and $|A_H^{\rm bare}|$ (dashed, red line) vs $m_{\gamma \gamma}$ for
$m_H = 750$~GeV. Right figure: the same amplitudes, but with the
stop-antistop propagator replaced with a Breit-Wigner resonance, as in
Eq.~(\ref{Breit-Wigner}).
}
\end{figure}

In Fig.~\ref{fig:8Mamp}, we show  $|A_{\rm tot}|$, $|A^{\rm
bare}_H|$, and $|A_{\tilde t\tilde t}^{\rm bare}|$ for $m_H = 750$~GeV. The left
and right panels show the results that are obtained in the
Coulomb-Schr\"odinger case and the Breit-Wigner case, respectively.
In both the Coulomb-Schr\"odinger case and the Breit-Wigner case,
there is a clear shift of the physical poles at maximal mixing away from
the stop-antistop threshold. However, there are several important
differences between the Coulomb-Schr\"odinger amplitude and the
Breit-Wigner amplitude. First, at maximal mixing, the larger-mass peak
that is present in the Breit-Wigner amplitude has almost disappeared in
the Coulomb-Schr\"odinger amplitude. The reason for this is that the
larger-mass physical state has a very large decay width into a
stop-antistop pair. The very narrow peak in the Coulomb-Schr\"odinger
amplitude near threshold does not correspond to the larger-mass peak in
the Breit-Wigner amplitude. Rather, it arises from the logarithmic term
in Eq.~(\ref{G-C-S}). It gives a small contribution to the cross
section. We also see that the width of the lower-mass peak is
significantly smaller in the Coulomb-Schr\"odinger amplitude than in the
Breit-Wigner amplitude, while its height is about the same. This
results in a reduced contribution of this peak to the cross section.

\begin{figure}
\epsfig{file=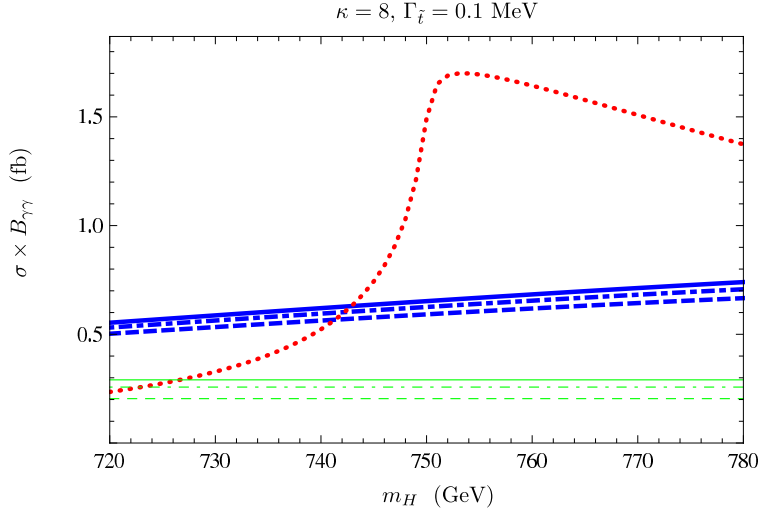,width=10cm}
\caption{\label{fig:8MCS}%
Cross sections for the case $\kappa=8$, $\Gamma_{\tilde t}=0.1$~MeV:
$\sigma_{\rm tot}$ (thick, blue lines),
$\sigma_H^{\rm bare} $ (dotted, red line), and $\sigma_{\tilde t
\tilde t}^{\rm bare}$ (thin, green lines) vs $m_H$. In the
cases of $\sigma_{\rm tot}$ and $\sigma_{\tilde t \tilde t}^{\rm
bare}$, the dashed, dashed-dotted, and solid lines correspond,
respectively, to taking $1$, $3$, or all terms in the sum in
Eq.~(\ref{G-C-Sa}).
}
\end{figure}

In Fig.~\ref{fig:8MCS}, we show $\sigma_{\rm tot}$, $\sigma_H^{\rm
bare}$, and $\sigma_{\tilde t \tilde t}^{\rm bare}$, as functions
of $m_H$. 
A comparison with the right panel of Fig.~\ref{fig:BR-8n} shows that the
shape of $\sigma_{\rm tot}$ in the Coulomb-Schr\"odinger case is similar
to the shape of $\sigma_{\rm tot}$ in the Breit-Wigner case: Both are
fairly featureless, although the slopes are different. As in the case
of the Breit-Wigner cross section, this featureless nature of the
Coulomb-Schr\"odinger cross section is a consequence of three properties
of the amplitude: (1) the short-distance coefficients are dominated by
their imaginary parts, and so the mixing of the short-distance
coefficients produces no pronounced peaks or dips; (2) the Higgs
short-distance coefficients dominate over the stop-antistop
short-distance coefficients, and so the stoponium peak does not
contribute greatly to the cross section; and (3) the height and width of the
lower-mass resonance are not very different from those of a Higgs
resonance over the range of $m_H$.

The cross section is considerably smaller in the Coulomb-Schr\"odinger
case than in the Breit-Wigner case. The reasons for this are the
disappearance of the higher-mass peak in the Coulomb-Schr\"odinger case,
owing to its large width into a stop-antistop pair, and the narrowing of
the lower-mass peak while its width remains constant. 

In the case $\kappa=8$, in comparison to the case of $\kappa = 1$,
$\sigma_H^{\rm bare}$ is greatly enhanced by the large
Higgs-stop-antistop coupling and is greater than $\sigma_{\tilde t
\tilde t}^{\rm bare}$. At $m_H = 750$~GeV, we see the threshold
enhancement of $\sigma_H^{\rm bare}$ that is associated with the  Higgs
diphoton and digluon form factors. However, $\sigma_{\rm tot}$ does not
show a similar threshold enhancement because of the shifts of the masses
of the physical states away from threshold. Although $\sigma_{\rm tot}$
is enhanced at small values of the Higgs mass with respect to 
$\sigma_H^{\rm bare}$, this enhancement becomes a suppression of factors
of a few at values of the Higgs mass that are close to the stop-antistop
production threshold.\footnote{We have also
examined the cross sections for the intermediate value $\kappa=5$. The
results are qualitatively similar to those at $\kappa=8$, except that
$\sigma_H^{\rm bare}$ is reduced relative to $\sigma_{\rm tot}$, and, so,
there is a mild enhancement of $\sigma_{\rm tot}$ relative to
$\sigma_H^{\rm bare}$.}

\subsection{$\kappa = 8$, $\Gamma_{\tilde t} = 0.1$~GeV}

Finally, we discuss the case of a large Higgs-stop-antistop coupling, 
$\kappa=8$, and a large stop width, $\Gamma_{\tilde t} = 0.1$~GeV.

\begin{figure}
\epsfig{file=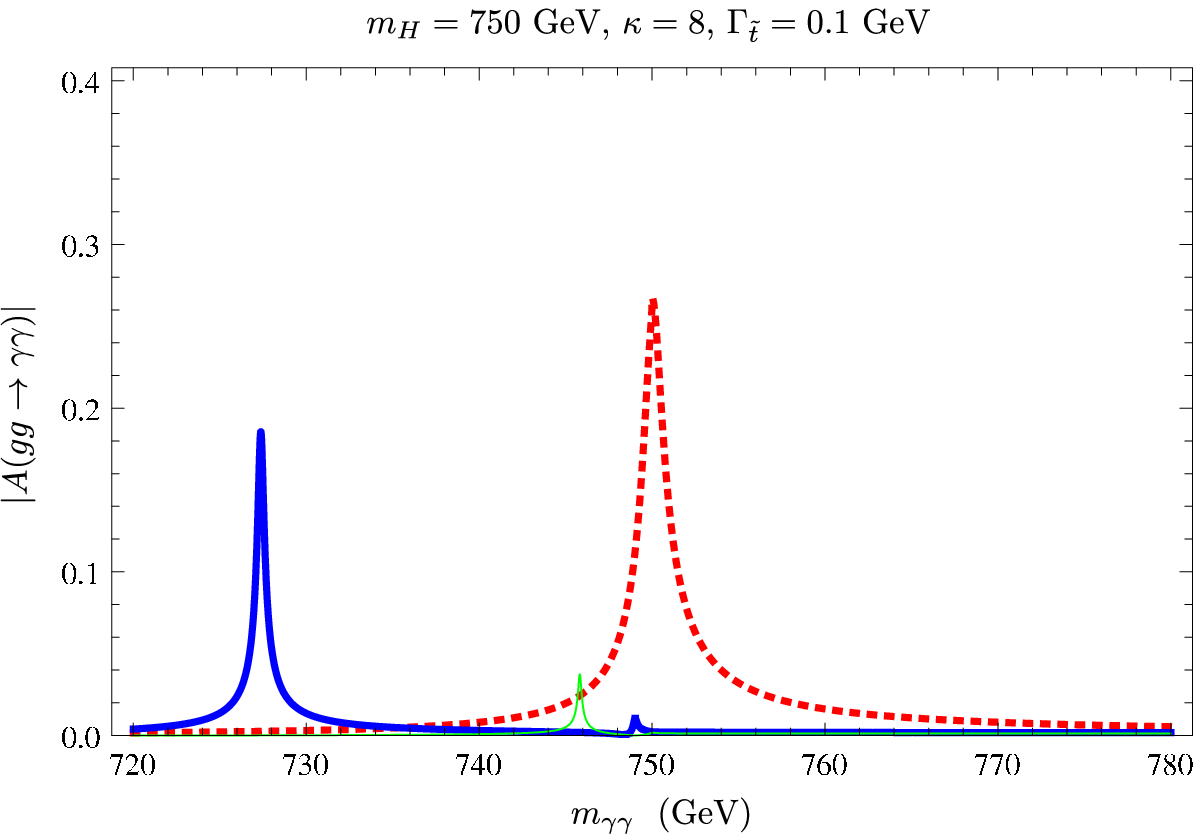,width=0.45\columnwidth} 
\epsfig{file=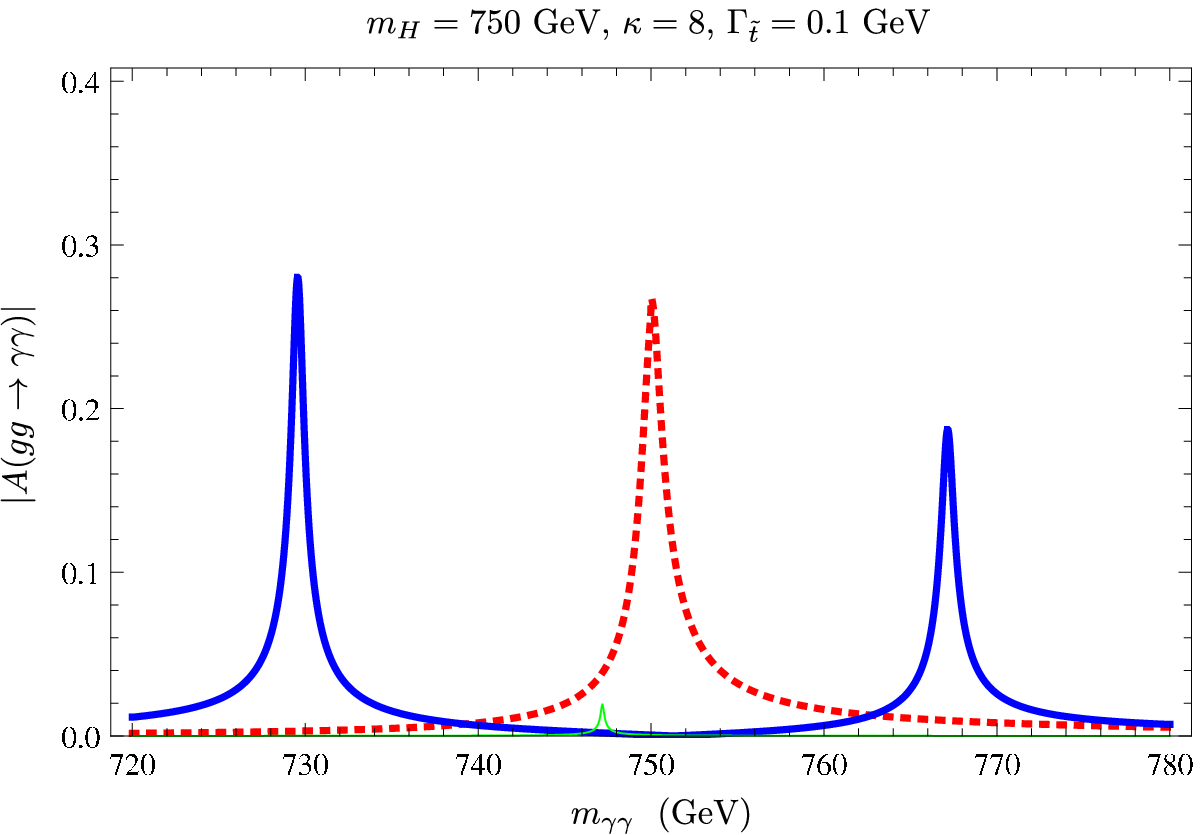,width=0.45\columnwidth}
\caption{\label{fig:8Gamp}%
Amplitudes for the case $\kappa=8$, $\Gamma_{\tilde t}=0.1$~GeV.
Left figure: the amplitudes $|A_{\rm tot}|$ (thick, blue line),
$|A{\tilde t\tilde t}^{\rm bare}|$ (thin, green line), 
and $|A_H^{\rm bare}|$ (dashed, red line) vs $m_{\gamma \gamma}$ for
$m_H = 750$~GeV. Right figure: the same amplitudes, but with the
stop-antistop propagator replaced with a Breit-Wigner resonance, as in
Eq.~(\ref{Breit-Wigner}).
}
\end{figure}

In Fig.~\ref{fig:8Gamp}, we show $|A_{\rm tot}|$, $|A^{\rm
bare}_H|$, and $|A_{\tilde t\tilde t}^{\rm bare}|$ for $m_H = 750$~GeV.
The left and right panels show the results that are obtained in the
Coulomb-Schr\"odinger case and the Breit-Wigner case, respectively. As
in the case of the smaller stop width, at maximal mixing there is a
clear shift of the physical poles away from the stop-antistop threshold.
There are again several important differences between the
Coulomb-Schr\"odinger amplitude and the Breit-Wigner amplitude. At
maximal mixing, the larger-mass peak that is present in the Breit-Wigner
amplitude has almost disappeared in the Coulomb-Schr\"odinger amplitude,
owing to the large decay width of the larger-mass physical state into a
stop-antistop pair. There is again a structure near threshold that
appears only in the Coulomb-Schr\"odinger amplitude  that arises from
the logarithmic term in Eq.~(\ref{G-C-S}), but in this large-stop-width
case, the structure has nearly disappeared. We also see that both the
height and the width of the lower-mass peak are significantly smaller in
the Coulomb-Schr\"odinger amplitude than in the Breit-Wigner amplitude.
These changes in the lower-mass peak result in a greatly reduced
contribution of the lower-mass peak to the cross section. In this
larger-stop-width case, the stoponium peak in $|A^{\rm bare}_H|$ is much
broader than in the smaller-stop-width case and is so small as to be
nearly invisible.

\begin{figure}
\epsfig{file=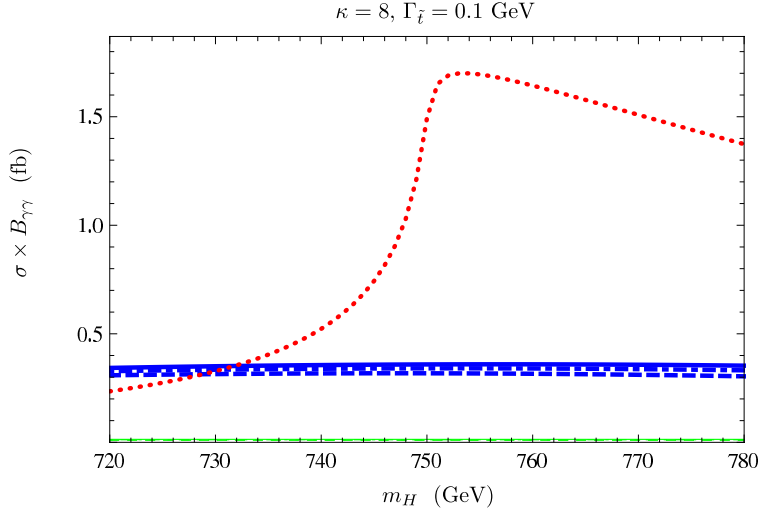,width=10cm}
\caption{\label{fig:8GCS}%
Cross sections for the case $\kappa=8$, $\Gamma_{\tilde t}=0.1$~GeV:
$\sigma_{\rm tot}$ (thick, blue lines),
$\sigma_H^{\rm bare} $ (dotted, red line), and $\sigma_{\tilde t
\tilde t}^{\rm bare}$ (thin, green lines) vs $m_H$. In the
cases of $\sigma_{\rm tot}$ and $\sigma_{\tilde t \tilde t}^{\rm
bare}$, the dashed, dashed-dotted, and solid lines correspond,
respectively, to taking $1$, $3$, or all terms in the sum in
Eq.~(\ref{G-C-Sa}).
}
\end{figure}

In Fig.~\ref{fig:8GCS}, we show $\sigma_{\rm tot}$, $\sigma_H^{\rm
bare}$, and $\sigma_{\tilde t \tilde t}^{\rm bare}$, as functions
of $m_H$.
A comparison with the right panel of Fig.~\ref{fig:BR-8w} shows that the
shape of $\sigma_{\rm tot}$ in the Coulomb-Schr\"odinger case is again
similar to the shape of $\sigma_{\rm tot}$ in the Breit-Wigner case.
The cross section is featureless for the reasons that we mentioned in
the discussion of the cross section for the narrower stop width.
For this larger stop-width, the cross section is much smaller in
the Coulomb-Schr\"odinger case than in the Breit-Wigner case. As in the
case of the smaller stop width, the higher-mass peak has disappeared in
the Coulomb-Schr\"odinger amplitude, owing to the large width of the
higher-mass peak into a stop-antistop pair. Furthermore, the reduction in
both the height and the width of the lower-mass peak has resulted in an
additional reduction of $\sigma_{\rm tot}$, relative to its values 
in the smaller-stop-width case.

Again, we see that the larger value of the Higgs-stop-antistop
coupling greatly enhances $\sigma_H^{\rm bare}$. Owing to the larger
stop width, $\sigma_{\tilde t \tilde t}^{\rm bare}$ has almost
disappeared in the figure. At $m_H = 750$~GeV, we again see the
threshold enhancement of $\sigma_H^{\rm bare}$ that is associated with
the  Higgs diphoton and digluon form factors. As in the
smaller-stop-width-case, $\sigma_{\rm tot}$ does not show a similar
threshold enhancement because of the shifts of the masses of the
physical states away from threshold. In this larger-stop-width case,
$\sigma_{\rm tot}$ is comparable to $\sigma_H^{\rm bare}$ at small
values of the Higgs mass and is much smaller than $\sigma_H^{\rm bare}$
at values of the Higgs mass that are close to the stop-antistop
production threshold.\footnote{At $\kappa=5$, the cross-section results are
qualitatively similar to those at $\kappa=8$, except that $\sigma_H^{\rm
bare}$ is reduced relative to $\sigma_{\rm tot}$, and, so, there is a mild
enhancement of $\sigma_{\rm tot}$ relative to $\sigma_H^{\rm bare}$,
except in a small region of $m_H$ between $750$ and $770$~GeV.}

\section{Conclusions\label{sec:conclusions}}

The system of a heavy Higgs boson that is coupled to a stop-antistop
pair exhibits some interesting field-theoretic phenomena near the
stop-antistop production threshold. This system has attracted
interest in the context of the production of a heavy Higgs boson near
the stop-antistop threshold and its subsequent decay to two photons
\cite{Djouadi:2016oey}, owing to the perturbative enhancement of the
Higgs couplings to photons and gluons near threshold. However, as is
well known, the appearance of Coulomb infrared singularities near
threshold invalidates the use of fixed-order perturbation theory. These
Coulomb singularities first appear at two-loop order in the
Higgs-to-diphoton and Higgs-to-digluon form factors. They occur when $v$
(one-half the stop-antistop relative velocity in the stop-antistop CM
frame) goes to zero, and they are a manifestation of the general
phenomenon of $1/v$ enhancements of two-particle amplitudes near
threshold. These enhancements require an all-orders treatment, and they
lead, among other things, to the formation of stoponium bound states.

A discussion of nonperturbative threshold effects is given in
Ref.~\cite{Djouadi:2016oey} and focuses on the enhancements of the
Higgs-to-diphoton and Higgs-to-digluon form factors that are induced by
the stop-antistop bound states~\cite{Melnikov:1994jb}. The discussion
in Ref.~\cite{Djouadi:2016oey} suggests that nonperturbative threshold
effects produce an enhancement of the digluon-to-diphoton cross section,
relative to the predictions for the cross section that are based on
perturbative treatments of the Higgs-to-diphoton and Higgs-to-digluon
form factors near threshold. However, a correct treatment of the
threshold effects also requires a complete analysis of
Higgs-stop-antistop mixing effects.

In this paper, we have formulated the calculation of the threshold
enhancements to the digluon-to-diphoton cross section in terms of
the scalar-quark analogues of the effective field theories NRQED and
NRQCD. Our treatment is valid up to corrections of relative order $v^2$.
The effective theory gives a complete accounting of Higgs-stop-antistop
mixing in the threshold region. We have studied these enhancement and
mixing effects numerically by making use of a model Green's function for
the stop-antistop system, namely, the Coulomb-Schr\"odinger Green's
function. The Coulomb-Schr\"odinger Green's function does not
correctly account for the QCD confining potential, which should be
Coulombic only for the lowest-lying stoponium bound states. Therefore,
we have considered the case in which the expression for the
Coulomb-Schr\"odinger Green's function is truncated  so that it
contains only a few bound states. This approach retains only bound
states for which the Coulombic approximation is expected to be valid,
and it is in keeping with the actual stoponium spectrum, which is
expected have only a few bound states. At a qualitative level, we have
checked that the results that we have obtained are independent of the
number of bound states that we have retained. Moreover, the quantitative
differences that are associated with the inclusion of heavier bounds
states are small, giving us confidence that our conclusions are not
dependent on the specifics of the model Green's function that we have
chosen.

We have also investigated a simplified model in which the
stop-antistop Green's function is represented by a simple Breit-Wigner
resonance. This simplified model exhibits some, but not all, of the
qualitative features of the more complicated Coulomb-Schr\"odinger
model.

We have found that the Higgs-stop-antistop mixing produces three general
effects that are very significant. First, the mixing leads to mass
eigenstates whose widths are larger than the widths of the stoponium
states. For a single stoponium state and for large values of the
Higgs-stop-antistop coupling, the widths of the mass eigenstates at
threshold approach the average of the Higgs and stoponium widths.
These increases in the widths, and the concomitant reductions in the
peak heights, reduce the contributions to the cross section relative
to the contribution that would be obtained from a narrow stoponium state.
Second, the physical masses are shifted from the input Higgs and
stoponium masses, and, when the Higgs mass is near threshold, the
physical masses are displaced away from the threshold region. This
effect is particularly important for large values of the
Higgs-stop-antistop coupling and can render the perturbative threshold
enhancements inoperative. Third, when the Higgs-stop-antistop coupling
is large, the displacement of the mass of the higher-mass physical state
to a point above threshold can give that state a very large width into a
stop-antistop pair, resulting in a drastic reduction of its contribution
to the cross section.

In addition, to these general effects, there are some effects that
depend on the details of the couplings of the Higgs boson and the
stop-antistop pair to photons and gluons and on the details of the
Coulomb-Schr\"odinger Green's function. For example, the couplings can
mix in such a way as to produce a peak near threshold that has
nothing to do with the threshold enhancements that are associated with
the perturbative Higgs-digluon and Higgs-diphoton form factors. The
Coulomb-Schr\"odinger Green's function can also lead to changes in the
heights and widths of the physical peaks in the amplitudes, relative
to their heights and widths in the simple Breit-Wigner model. These
effects are driven largely by the term of lowest order in
$\alpha_s$, in the Coulomb-Schr\"odinger Green's function. That term is
universal in that it is independent of the nature of the
squark-antiquark static potential. However, the details of the effects
that arise from it  seem to depend on nonuniversal features of the
Coulomb-Schr\"odinger Green's function.

In general, for large values of the heavy-Higgs coupling to the
stop-antistop pair, the mixing effects result in suppressions of the
digluon-to-diphoton cross section at threshold relative to the cross
section that is predicted in one-loop perturbation theory. The precise
suppression factor depends not only on the Higgs-stop-antistop coupling 
but also on the stop width. We remind the reader that, because our
focus is on the formulation of the calculation and on the qualitative
features of the threshold physics, we have computed the Higgs couplings
to digluons and diphotons at the one-loop level, and, so, one should
take care in comparing our numerical results with those in the
literature, which often include two-loop effects. 

Although we have concentrated on the case of the Higgs-stop-antistop
interaction, the theoretical framework that we have developed is
applicable to the coupling of Higgs bosons to other scalar particles in
the region near the particle-antiparticle threshold. It can also be
generalized easily to the case of a Higgs boson coupled to heavy
fermions and to calculations of rates to different final states. For
example, one could study the case of a $\tau^+ \tau^-$ final
state by replacing the $\gamma\gamma$ short-distance coefficients in
Eq.~(\ref{matrix-amps-a}) with the corresponding $\tau^+ \tau^-$
short-distance coefficients.\footnote{We note that
$C_{\tau^+ \tau^- \tilde t\tilde t}$ vanishes if one neglects
electromagnetic and weak interactions.} We reserve the study of these
additional cases for a separate publication.

\appendix
\section{Diagonal form of the amplitude in the general case 
\label{sec:C-S-diagonal}}

In the general case, which includes the example of the
Coulomb-Schr\"odinger Green's function, we can write
Eq.~(\ref{matrix-amps-a}) as
\begin{eqnarray}
A_{\rm tot} (gg \to \gamma\gamma)
&=&\left(
\begin{array}{cc}
C_{ggH} & \hat{C}_{gg\tilde t\tilde t}
\end{array}
\right)
\left(
\begin{array}{cc}
S_{H}^{-1}(\hat s) & \;\;\;\; -\hat{C}_{H\tilde t\tilde t}\\
-\hat{C}_{H\tilde t\tilde t} & \;\;\;\;\;\; \hat{G}_{\tilde t\tilde t}^{-1}(\hat s)
\end{array}
\right)^{-1}
\left(
\begin{array}{c}                                                        
C_{\gamma\gamma H}\\ 
\hat{C}_{\gamma\gamma\tilde t\tilde t}             
\end{array}
\right),
\label{matrix-amps-for-diagonalization}
\end{eqnarray}
where $\hat{G}_{\tilde t\tilde t}(\hat{s})=
\tilde{G}_{\tilde t\tilde t}(\hat{s})/N_{\tilde t\tilde t}^2$, with 
$N_{\tilde t\tilde t}^2$ given in Eq.~(\ref{Ntt}).\footnote{The choice 
of $N_{\tilde t\tilde t}^2$ is somewhat arbitrary. Here, we have chosen 
$N_{\tilde t\tilde t}^2$ so as to be consistent with the choice that we 
made in the Breit-Wigner case.}
The eigenvalues of the matrix in
Eq.~(\ref{matrix-amps-for-diagonalization}) whose inverse is taken are
given by
\begin{equation}
\alpha_\pm(\hat{s})=-\frac{i}{2}\left\{
\frac{1}{-iS_H(\hat{s})}+\frac{1}{-i\tilde{G}_{\tilde t\tilde t}(\hat{s})}
\pm\sqrt{
\biggl[\frac{1}{-iS_H(\hat{s})}
-\frac{1}{-i\tilde{G}_{\tilde t\tilde t}(\hat{s})}
\biggr]^2+4|C_{H\tilde t\tilde t}|^2
}
\right\},
\end{equation}
and the tangent of the rotation angle of the similarity transformation
that diagonalizes that matrix is given by
\begin{eqnarray}
\tan[\theta(\hat{s})]&=&\frac{2|\hat{C}_{H\tilde t\tilde t}|}
{\displaystyle \sqrt{\biggl[\frac{1}{-iS_H(\hat{s})}
-\frac{1}{-i\tilde{G}_{\tilde t\tilde t}(\hat{s})}
\biggr]^2
+4|\hat{C}_{H\tilde t\tilde t}|^2}
+ \bigg [ 
\frac{1}{-iS_H(\hat{s})}-\frac{1}{-i\tilde{G}_{\tilde t\tilde t}(\hat{s})
}
\bigg ] 
}.
\label{similarity-angle-general}
\end{eqnarray}
We see that both the eigenvalues and the rotation angle now depend on 
$\hat{s}$.

The physical-state poles are located at the values $\hat{s}=\hat{s}_\pm$
for which $\alpha_\pm(\hat{s})$ vanishes. [Note that there may be more
than one value of $\hat{s}_\pm$ for which $\alpha_\pm(\hat{s})$
vanishes.] Near a pole, the eigenvalues of the inverse matrix that
appears in Eq.~(\ref{matrix-amps-for-diagonalization}) are
\begin{subequations}%
\begin{equation}
\frac{i}{Z_\pm^{-1}(\hat{s}-m_\pm^2)+iI_\pm}
=\frac{iZ_\pm}{\hat{s}-m_\pm^2 +i m_\pm\Gamma_\pm},
\label{C-S-pole}
\end{equation}
where
\begin{eqnarray}
m_\pm^2&=&{\rm Re}(\hat{s}_\pm),\\
I_{\pm}&=&{\rm Im}[i \alpha_\pm(\hat{s})]|_{\hat{s}=m_\pm^2},\\
Z_\pm^{-1}&=&\frac{\partial}{\partial \hat{s}}
{\rm Re}[i \alpha_\pm(\hat{s})]|_{\hat{s}=m_\pm^2},\\
m_\pm \Gamma_\pm&=&I_\pm Z_\pm,
\end{eqnarray}
and the tangent of the rotation angle is given by 
\begin{equation}
\tan[\theta(m_\pm^2)]=\tan[\theta(\hat{s})]|_{\hat{s}=m_\pm}.
\end{equation}
\end{subequations}%

\begin{acknowledgments}                                      

G.T.B.\ and H.S.C.\ would like to thank Estia Eichten for a helpful discussion.
C.E.M.W.\ would like to thank Marcela Carena, Abdelhak Djouadi,
Ahmed Ismail, Ian Low, Steve Martin, and Nausheen Shah for useful discussions.
The work of G.T.B., H.S.C., and C.E.M.W.\
is supported by the U.S.\ Department of Energy, Division of High Energy
Physics, under Contract No. DE-AC02-06CH11357. The submitted
manuscript has been created in part by UChicago Argonne, LLC, Operator
of Argonne National Laboratory. Argonne, a U.S.\ Department of Energy
Office of Science laboratory, is operated under Contract No.
DE-AC02-06CH11357. The U.S. Government retains for itself, and others
acting on its behalf, a paid-up nonexclusive, irrevocable worldwide
license in said article to reproduce, prepare derivative works,
distribute copies to the public, and perform publicly and display
publicly, by or on behalf of the Government. The work of C.E.M.W. at the
University of Chicago is partially supported by the U.S. Department of 
Energy, under Contract No. DE-SC0009924.
The work of H.S.C. at CERN is partially supported by the 
Korean Research Foundation through the CERN-Korea fellowship program.

\end{acknowledgments}



\begin{thebibliography}{}

\bibitem{Haber:1984rc} 
  H.~E.~Haber and G.~L.~Kane,
  Phys.\ Rept.\  {\bf 117}, 75 (1985).
  doi:10.1016/0370-1573(85)90051-1
  
\bibitem{Nilles:1983ge} 
  H.~P.~Nilles,
  Phys.\ Rept.\  {\bf 110}, 1 (1984).
  doi:10.1016/0370-1573(84)90008-5

\bibitem{Drees:2004jm} 
  M.~Drees, R.~Godbole and P.~Roy,
  {\it Theory and Phenomenology of Sparticles} 
  (World Scientific, Singapore, 2004)
  
\bibitem{Baer:2006rs} 
  H.~Baer and X.~Tata, {\it 
  Weak scale supersymmetry: From superfields to scattering events}, 
  (Cambridge University Press, Cambridge, England, 2006).



\bibitem{Martin:1997ns} 
  S.~P.~Martin,
  Adv.\ Ser.\ Direct.\ High Energy Phys.\  {\bf 21}, 1 (2010)
  [Adv.\ Ser.\ Direct.\ High Energy Phys.\  {\bf 18}, 1 (1998)]
  doi:$10.1142/9789812839657_0001, 10.1142/9789814307505_0001$
  [hep-ph/9709356].



  


\bibitem{Drees:1993uw} 
  M.~Drees and M.~M.~Nojiri,
  Phys.\ Rev.\ D {\bf 49}, 4595 (1994)
  doi:10.1103/PhysRevD.49.4595
  [hep-ph/9312213].


\bibitem{Martin:2008sv} 
  S.~P.~Martin,
  Phys.\ Rev.\ D {\bf 77}, 075002 (2008)
  doi:10.1103/PhysRevD.77.075002
  [arXiv:0801.0237 [hep-ph]].



\bibitem{Martin:2009dj} 
  S.~P.~Martin and J.~E.~Younkin,
  Phys.\ Rev.\ D {\bf 80}, 035026 (2009)
  doi:10.1103/PhysRevD.80.035026
  [arXiv:0901.4318 [hep-ph]].



\bibitem{Younkin:2009zn} 
  J.~E.~Younkin and S.~P.~Martin,
  Phys.\ Rev.\ D {\bf 81}, 055006 (2010)
  doi:10.1103/PhysRevD.81.055006
  [arXiv:0912.4813 [hep-ph]].

\bibitem{Kats:2016kuz} 
  Y.~Kats and M.~J.~Strassler,
  JHEP {\bf 1605}, 092 (2016)
  doi:10.1007/JHEP05(2016)092
  [arXiv:1602.08819 [hep-ph]].
  
  
\bibitem{Choudhury:2016jbc} 
  D.~Choudhury and K.~Ghosh,
  arXiv:1605.00013 [hep-ph].

  
\bibitem{Carena:2016bnq} 
  M.~Carena, P.~Huang, A.~Ismail, I.~Low, N.~R.~Shah and C.~E.~M.~Wagner,
  Phys.\ Rev.\ D {\bf 94}, no. 11, 115001 (2016)
  doi:10.1103/PhysRevD.94.115001
  [arXiv:1606.06733 [hep-ph]].


\bibitem{Branco:2011iw} 
  G.~C.~Branco, P.~M.~Ferreira, L.~Lavoura, M.~N.~Rebelo, M.~Sher and J.~P.~Silva,
  Phys.\ Rept.\  {\bf 516}, 1 (2012)
  doi:10.1016/j.physrep.2012.02.002
  [arXiv:1106.0034 [hep-ph]].
  
 


\bibitem{Drees:1989du} 
  M.~Drees and K.~i.~Hikasa,
  Phys.\ Rev.\ D {\bf 41}, 1547 (1990).
  doi:10.1103/PhysRevD.41.1547
  
  

\bibitem{Djouadi:2016oey} 
  A.~Djouadi and A.~Pilaftsis,
  arXiv:1605.01040 [hep-ph].


\bibitem{Melnikov:1994jb} 
  K.~Melnikov, M.~Spira and O.~I.~Yakovlev,
  Z.\ Phys.\ C {\bf 64}, 401 (1994)
  doi:10.1007/BF01560100
  [hep-ph/9405301].



\bibitem{Caswell:1985ui} 
  W.~E.~Caswell and G.~P.~Lepage,
  Phys.\ Lett.\ B {\bf 167}, 437 (1986).
  doi:10.1016/0370-2693(86)91297-9

\bibitem{Lepage:1992tx} 
  G.~P.~Lepage, L.~Magnea, C.~Nakhleh, U.~Magnea and K.~Hornbostel,
  Phys.\ Rev.\ D {\bf 46}, 4052 (1992)
  doi:10.1103/PhysRevD.46.4052
  [hep-lat/9205007].

\bibitem{Bodwin:1994jh} 
  G.~T.~Bodwin, E.~Braaten and G.~P.~Lepage,
  Phys.\ Rev.\ D {\bf 51}, 1125 (1995)
  Erratum: [Phys.\ Rev.\ D {\bf 55}, 5853 (1997)]
  doi:10.1103/PhysRevD.55.5853, 10.1103/PhysRevD.51.1125
  [hep-ph/9407339].

\bibitem{ATLAS:2016jcu} 
  The ATLAS collaboration [ATLAS Collaboration],
  ATLAS-CONF-2016-029.

\bibitem{CMS:2016owr} 
  CMS Collaboration [CMS Collaboration],
  CMS-PAS-EXO-16-018.


\bibitem{Strumia:2016wys} 
  A.~Strumia,
  arXiv:1605.09401 [hep-ph].

\bibitem{Casas:1995pd} 
  J.~A.~Casas, A.~Lleyda and C.~Munoz,
  Nucl.\ Phys.\ B {\bf 471}, 3 (1996)
  doi:10.1016/0550-3213(96)00194-0
  [hep-ph/9507294].
  
\bibitem{Blinov:2013fta} 
  N.~Blinov and D.~E.~Morrissey,
  JHEP {\bf 1403}, 106 (2014)
  doi:10.1007/JHEP03(2014)106
  [arXiv:1310.4174 [hep-ph]].
  
\bibitem{Chowdhury:2013dka} 
  D.~Chowdhury, R.~M.~Godbole, K.~A.~Mohan and S.~K.~Vempati,
  JHEP {\bf 1402}, 110 (2014)
  doi:10.1007/JHEP02(2014)110
  [arXiv:1310.1932 [hep-ph]].

\bibitem{Kim:2014yaa} 
  C.~Kim, A.~Idilbi, T.~Mehen and Y.~W.~Yoon,
  Phys.\ Rev.\ D {\bf 89}, no. 7, 075010 (2014)
  doi:10.1103/PhysRevD.89.075010
  [arXiv:1401.1284 [hep-ph]].

\bibitem{Bauer:2000yr} 
  C.~W.~Bauer, S.~Fleming, D.~Pirjol and I.~W.~Stewart,
  Phys.\ Rev.\ D {\bf 63}, 114020 (2001)
  doi:10.1103/PhysRevD.63.114020
  [hep-ph/0011336].

\bibitem{Spira:1995rr} 
  M.~Spira, A.~Djouadi, D.~Graudenz and P.~M.~Zerwas,
  Nucl.\ Phys.\ B {\bf 453}, 17 (1995)
  doi:10.1016/0550-3213(95)00379-7
  [hep-ph/9504378].
\bibitem{Harlander:2012pb} 
  R.~V.~Harlander, S.~Liebler and H.~Mantler,
  Comput.\ Phys.\ Commun.\  {\bf 184}, 1605 (2013)
  doi:10.1016/j.cpc.2013.02.006
  [arXiv:1212.3249 [hep-ph]].
\bibitem{Bagnaschi:2014zla} 
  E.~Bagnaschi, R.~V.~Harlander, S.~Liebler, H.~Mantler, P.~Slavich and A.~Vicini,
  JHEP {\bf 1406}, 167 (2014)
  doi:10.1007/JHEP06(2014)167
  [arXiv:1404.0327 [hep-ph]].
\bibitem{Dittmaier:2014sva} 
  S.~Dittmaier, P.~H\"afliger, M.~Kr\"amer, M.~Spira and M.~Walser,
  Phys.\ Rev.\ D {\bf 90}, no. 3, 035010 (2014)
  doi:10.1103/PhysRevD.90.035010
  [arXiv:1406.5307 [hep-ph]].


  
\bibitem{Wilczek:1977zn}
  F.~Wilczek,
  Phys.\ Rev.\ Lett.\  {\bf 39} (1977) 1304.
  doi:10.1103/PhysRevLett.39.1304
  
\bibitem{Ellis:1979jy} 
  J.~R.~Ellis, M.~K.~Gaillard, D.~V.~Nanopoulos and C.~T.~Sachrajda,
  Phys.\ Lett.\ B {\bf 83}, 339 (1979).
  doi:10.1016/0370-2693(79)91122-5
  
\bibitem{Gunion:1989we} 
  J.~F.~Gunion, H.~E.~Haber, G.~L.~Kane and S.~Dawson,
  Front.\ Phys.\  {\bf 80}, 1 (2000).
  
  
\bibitem{Ellis:1975ap} 
  J.~R.~Ellis, M.~K.~Gaillard and D.~V.~Nanopoulos,
  Nucl.\ Phys.\ B {\bf 106}, 292 (1976).
  doi:10.1016/0550-3213(76)90382-5
  
\bibitem{Shifman:1979eb} 
  M.~A.~Shifman, A.~I.~Vainshtein, M.~B.~Voloshin and V.~I.~Zakharov,
  Sov.\ J.\ Nucl.\ Phys.\  {\bf 30}, 711 (1979)
  [Yad.\ Fiz.\  {\bf 30}, 1368 (1979)].
  
\bibitem{Djouadi:1993ji} 
  A.~Djouadi, M.~Spira and P.~M.~Zerwas,
  Phys.\ Lett.\ B {\bf 311}, 255 (1993)
  doi:10.1016/0370-2693(93)90564-X
  [hep-ph/9305335].
  


\bibitem{Brambilla:1999xf} 
  N.~Brambilla, A.~Pineda, J.~Soto and A.~Vairo,
  Nucl.\ Phys.\ B {\bf 566}, 275 (2000)
  doi:10.1016/S0550-3213(99)00693-8
  [hep-ph/9907240].

\bibitem{Dreiner:2016wwk} 
  H.~K.~Dreiner, M.~E.~Krauss, B.~O'Leary, T.~Opferkuch and F.~Staub,
  Phys.\ Rev.\ D {\bf 94}, no. 5, 055013 (2016)
  doi:10.1103/PhysRevD.94.055013
  [arXiv:1606.08811 [hep-ph]].


\bibitem{Eichten:1974af} 
  E.~Eichten, K.~Gottfried, T.~Kinoshita, J.~B.~Kogut, K.~D.~Lane and T.~M.~Yan,
  Phys.\ Rev.\ Lett.\  {\bf 34}, 369 (1975)
  Erratum: [Phys.\ Rev.\ Lett.\  {\bf 36}, 1276 (1976)].
  doi:10.1103/PhysRevLett.34.369

 

\bibitem{Melnikov:1998pr}                                     
  K.~Melnikov and A.~Yelkhovsky,                                   
  Nucl.\ Phys.\ B {\bf 528}, 59 (1998)                                   
  doi:10.1016/S0550-3213(98)00348-4                                      
  [hep-ph/9802379].

\bibitem{Kiyo:2010jm}
  Y.~Kiyo, A.~Pineda and A.~Signer,
  Nucl.\ Phys.\ B {\bf 841}, 231 (2010)
  doi:10.1016/j.nuclphysb.2010.08.007
  [arXiv:1006.2685 [hep-ph]].

\bibitem{Bali:2000vr} 
  G.~S.~Bali {\it et al.} [SESAM and T$\chi$L Collaborations],
  Phys.\ Rev.\ D {\bf 62}, 054503 (2000)
  doi:10.1103/PhysRevD.62.054503
  [hep-lat/0003012].

\bibitem{Kim:2015zqa} 
  S.~Kim,
  Phys.\ Rev.\ D {\bf 92}, no. 9, 094505 (2015)
  doi:10.1103/PhysRevD.92.094505
  [arXiv:1508.07080 [hep-lat]].

\bibitem{Hagiwara:1990sq} 
  K.~Hagiwara, K.~Kato, A.~D.~Martin and C.~K.~Ng,
  Nucl.\ Phys.\ B {\bf 344}, 1 (1990).
  doi:10.1016/0550-3213(90)90683-5

\bibitem{Baumgart:2012pj} 
  M.~Baumgart and A.~Katz,
  JHEP {\bf 1208}, 133 (2012)
  doi:10.1007/JHEP08(2012)133
  [arXiv:1204.6032 [hep-ph]].

\bibitem{Pumplin:2002vw} 
  J.~Pumplin, D.~R.~Stump, J.~Huston, H.~L.~Lai, P.~M.~Nadolsky and W.~K.~Tung,
  JHEP {\bf 0207}, 012 (2002)
  doi:10.1088/1126-6708/2002/07/012
  [hep-ph/0201195].

\bibitem{Allanach:2015blv} 
  B.~C.~Allanach, P.~S.~B.~Dev and K.~Sakurai,
  Phys.\ Rev.\ D {\bf 93}, no. 3, 035010 (2016)
  doi:10.1103/PhysRevD.93.035010
  [arXiv:1511.01483 [hep-ph]].
 
\bibitem{Braaten:1980yq} 
  E.~Braaten and J.~P.~Leveille,
  Phys.\ Rev.\ D {\bf 22}, 715 (1980).
  doi:10.1103/PhysRevD.22.715

\bibitem{Drees:1990dq} 
  M.~Drees and K.~i.~Hikasa,
  Phys.\ Lett.\ B {\bf 240}, 455 (1990)
  Erratum: [Phys.\ Lett.\ B {\bf 262}, 497 (1991)].
  doi:10.1016/0370-2693(90)91130-4
  
 
\end{thebibliography}
\end{document}